\documentclass[superscriptaddress, reprint, amsmath, amssymb, aps, pra, floatfix]{revtex4-2}

\usepackage{bm}

\usepackage{mathtools}
\usepackage{amsfonts}
\usepackage{amssymb}
\usepackage{dsfont}
\usepackage{amsthm}
\usepackage[english]{babel}
\usepackage{graphicx}
\usepackage{amsmath}
\usepackage{algorithm,algorithmic}
\usepackage{lipsum}
\usepackage{wrapfig}
\usepackage{float}
\usepackage{enumitem} 
\usepackage{color} 
\usepackage{hyperref}
\usepackage[normalem]{ulem}
\usepackage{url}
\usepackage{appendix}
\usepackage{changes}
\definecolor{blue-violet}{rgb}{0.54, 0.17, 0.89}

\useunder{\uline}{\ul}{}

\hypersetup{colorlinks = true}

\newcommand{\norm}[1]{\left\lVert #1 \right\rVert}

\definecolor{blue(ryb)}{rgb}{0.01, 0.28, 1.0}

\newcommand{\Ep}{E_{\mathrm{p}}}

\newcommand{\Epb}{E_{\mathrm{p}, (b)}}
\newcommand{\Ept}{E_{\mathrm{p}, t}}
\newcommand{\Eptt}{E_{\mathrm{p}, t + 2\Delta t}}
\newcommand{\Hd}{H_{\mathrm{d}}}
\newcommand{\Hdj}{H_{\mathrm{d}, j}}
\newcommand{\Hp}{H_{\mathrm{p}}}

\newcommand{\Hpb}{H_{\mathrm{p}, (b)}}
\newcommand{\nd}{n_{\mathrm{d}}}
\newcommand{\np}{n_{\mathrm{p}}}
\newcommand{\Ud}{U_{\mathrm{d}}}
\newcommand{\Up}{U_{\mathrm{p}}}
\newcommand{\Uq}{U_{\mathrm{QAOA}}}

\usepackage{qcircuit}

\begin{document}

\title{Lyapunov control-inspired strategies for quantum combinatorial optimization}

\author{Alicia B. Magann} 

\affiliation{Quantum Algorithms and Applications Collaboratory, Sandia National Laboratories, Livermore, California 94550, USA}
\affiliation{Quantum Algorithms and Applications Collaboratory, Sandia National Laboratories, Albuquerque, New Mexico 87185, USA}
\affiliation{Department of Chemical \& Biological Engineering, Princeton University, Princeton, New Jersey 08544, USA}

\author{Kenneth M. Rudinger}
\affiliation{Quantum Algorithms and Applications Collaboratory, Sandia National Laboratories, Albuquerque, New Mexico 87185, USA}

\author{Matthew D. Grace}
\affiliation{Quantum Algorithms and Applications Collaboratory, Sandia National Laboratories, Livermore, California 94550, USA}

\author{Mohan Sarovar}
\affiliation{Quantum Algorithms and Applications Collaboratory, Sandia National Laboratories, Livermore, California 94550, USA}

\date{\today}

\begin{abstract}

The prospect of using quantum computers to solve combinatorial optimization problems via the quantum approximate optimization algorithm (QAOA) has attracted considerable interest in recent years. However, a key limitation associated with QAOA is the need to classically optimize over a set of quantum circuit parameters. This classical optimization can have significant associated costs and challenges. Here, we provide an expanded description of Lyapunov control-inspired strategies for quantum optimization, as presented in [Magann \emph{et al.}, \href{https://doi.org/10.1103/PhysRevLett.129.250502}{Phys. Rev. Lett. 129, 250502 (2022)}], that do not require any classical optimization effort. Instead, these strategies utilize feedback from qubit measurements to assign values to the quantum circuit parameters in a deterministic manner, such that the combinatorial optimization problem solution improves monotonically with the quantum circuit depth. Numerical analyses are presented that investigate the utility of these strategies towards MaxCut on weighted and unweighted 3-regular graphs, both in ideal implementations and also in the presence of measurement noise. We also discuss how how these strategies compare with QAOA, how they may be used to seed QAOA optimizations in order to improve performance for near-term applications, and explore connections to quantum annealing. 

\end{abstract}

\maketitle

\section{Introduction} 

Combinatorial optimization problems have a variety of broad and high-value applications, including in routing and scheduling problems \cite{golden2008vehicle,blazewicz1996job}. The desire to use quantum resources to aid in solving them has a long history, spanning the development of adiabatic and annealing-based strategies \cite{finnila1994quantum,PhysRevE.58.5355,brooke1999quantum}, as well as the development of early quantum algorithms \cite{durr1996quantum,durr2006quantum}. 
More recently, the quantum approximate optimization algorithm (QAOA) \cite{2014arXiv1411.4028F} was proposed in 2014, as a method for leveraging quantum computers to solve combinatorial optimization problems. 
In particular, QAOA is a method for determining an approximate solution to a combinatorial optimization problem by using a hybrid quantum-classical framework; a classical computer is utilized to iteratively minimize the value of the cost function, and the cost function is evaluated on a quantum computer using a parameterized quantum circuit. Since its development, QAOA has captured the attention of numerous theoretical and experimental groups, e.g.,  \cite{otterbach2017unsupervised, willsch2020benchmarking, abrams2019implementation, bengtsson2019quantum, Harrigan2021, Pagano2020}, particularly as a potential application of noisy, intermediate-scale quantum (NISQ) \cite{preskill_quantum_2018} devices. 

Recently, numerous connections have been made between QAOA and quantum optimal control (QOC) \cite{PRXQuantum.2.010101}, which is a strategy for identifying the controls needed to steer the dynamics of a quantum system in a desired manner by iteratively optimizing over a set of control functions or parameters \cite{Brif2011,Glaser2015}. Certain connections have rested on the control-theoretic notion of controllability, which implies that QOC solutions can be found for driving the dynamics of a system under consideration towards arbitrary objectives \cite{altafini2002controllability, albertini2002lie, schirmer2001complete, fu2001complete, turinici2003wavefunction, burgarth2013zero,  arenz2016universal,arenz2018controlling}. For instance, controllability considerations have recently been applied to show that QAOA can be computationally universal in certain circumstances \cite{2018arXiv181211075L, 2019arXiv190903123M}, and to assess the number of QAOA quantum circuit parameters needed to achieve controllability \cite{2019arXiv190608948B}.

A key challenge in identifying QAOA and QOC solutions is the difficulty of searching for the optimal QAOA and QOC parameters, respectively. If the number of QAOA layers, and correspondingly variational parameters, can be limited to $O(\mathrm{poly}(\log(n)))$, then the complexity scaling of the optimization can be polynomial in the number of qubits, $n$. However, this scaling belies the difficulty of such optimizations -- the fact is that QAOA, and most variational algorithms, require optimization over non-linear, stochastic cost functions with derivative information that is noisy and hard to obtain. Scaling classical optimization to thousands of parameters is challenging in this context, \emph{e.g.,} \cite{doi:10.1137/18M1177718}.

Even in the absence of noise, the difficulty of identifying the optimal parameters is determined in large part by the structure of the optimization landscape, and landscape features such as local optima, saddle points, and barren plateaus can complicate and hinder the optimization process \cite{chakrabarti2007quantum, russell2017control, mcclean2018barren, PhysRevA.102.042207,  PRXQuantum.1.020319, bittel2021training, lee2021towards,larocca2021diagnosing}. However, we note that a variety of alternate quantum control frameworks, including quantum tracking control \cite{Gross1993, Chen1995,PhysRevLett.118.083201, PhysRevA.101.053408, PhysRevLett.124.183201}, quantum Lyapunov control \cite{PhysRevLett.69.2172, doi:10.1063/1.467132, SUGAWARA1995113, OHTSUKI1998627, A902103E, doi:10.1063/1.1559680, 1272601, Mirrahimi2005ReferenceTT, doi:https://doi.org/10.1002/9780470431917.ch2}, and quantum feedback control \cite{doherty2000quantum, Wiseman_Milburn_2009, Combes_Kerckhoff_Sarovar_2017, zhang2017quantum}, have been developed that do not rely on the iterative classical optimization procedure inherent to QOC and thus do not share these challenges.

Here, we explore a new connection between quantum algorithms and quantum control theory, and develop strategies for quantum combinatorial optimization inspired by the theory of \emph{quantum Lyapunov control} (QLC). In particular, this article provides an expanded discussion of the details of these strategies, beyond that contained in \cite{magann2021feedbackbased}. Importantly, these QLC-inspired strategies do not involve any classical optimization. Instead, they use measurement-based feedback  to assign values to the quantum circuit parameters. We show that this feedback-based procedure yields a monotonically improving solution to the original combinatorial optimization problem, with respect to the depth of the quantum circuit. 

The remainder of this article is organized as follows. We begin by providing background on QAOA, as motivation for this work. This is followed by a description of certain aspects of QLC. We then introduce a Feedback-based ALgorithm for Quantum OptimizatioN (FALQON) inspired by QAOA and QLC, and discuss extensions that can be used to improve performance, including the addition of a reference perturbation, the implementation of an iterative procedure, and the introduction of additional control functions. We then discuss applications of FALQON towards solving the MaxCut problem. To this end, we provide numerical illustrations of the ideal performance of FALQON towards solving MaxCut on 3-regular graphs, and we also explore how it performs in the presence of measurement noise, and how its performance compares to that of QAOA. We go on to discuss how FALQON can be used to boost the performance of QAOA in NISQ applications, and explore connections to quantum annealing. We conclude with an outlook.

\section{Quantum approximate optimization algorithm}

Combinatorial optimization problems are concerned with identifying configurations of discrete optimization variables that best achieve one or multiple goals, as quantified by an associated cost function $C$. The quantum approximate optimization algorithm (QAOA) \cite{2014arXiv1411.4028F} is an approach for finding or approximating solutions to combinatorial optimization problems using quantum computers. It operates by first encoding the cost function $C$ into an Ising Hamiltonian, $\Hp$, which is diagonal in the quantum computational basis such that each eigenstate of $\Hp$ corresponds to a single spin configuration, which encodes a configuration of the associated optimization variables \cite{lucas2014ising}. The encoding is done such that the best solution to the combinatorial optimization problem is encoded in the ground state of $\Hp$. 

Using this encoding, QAOA is a hybrid quantum-classical algorithm for solving
\begin{equation}
   \min_{\{\gamma_k\},\{\beta_k\}} \langle \psi (\{\gamma_k\},\{\beta_k\})|\Hp|\psi(\{\gamma_k\},\{\beta_k\})\rangle\,.
\end{equation}
To do so, a classical computer is utilized to iteratively minimize the value of the objective function, evaluated as $\langle \psi (\{\gamma_k\},\{\beta_k\})|\Hp|\psi(\{\gamma_k\},\{\beta_k\})\rangle$, and optimized over the set of $2\ell$ parameters $\{\gamma_k\}_{k=1}^\ell $ and $\{\beta_k\}_{k=1}^\ell$.

The objective function value is determined at each iteration using a quantum computer, which prepares the multiqubit state $|\psi(\{\gamma_k\},\{\beta_k\})\rangle$ using a parameterized quantum circuit of the form 
\begin{equation}
    \Uq = \Ud(\beta_\ell) \Up(\gamma_\ell) \cdots \Ud(\beta_1) \Up(\gamma_1)\,,
    \label{eq:QAOA-sequence}
\end{equation}
such that $|\psi(\{\gamma_k\},\{\beta_k\})\rangle = \Uq |\psi_0\rangle$ for an initial state $|\psi_0\rangle$. 
The elements of the QAOA circuit, $\Up(\cdot)$ and $\Ud(\cdot)$, are created by simulating an evolution under $\Hp$ and under a \emph{driver} Hamiltonian $\Hd$, which is chosen to not commute with $\Hp$. Following the implementation of $\Uq$, measurements of $\Hp$ in the multiqubit state $|\psi(\{\gamma_k\},\{\beta_k\})\rangle$ allow for estimating the value of the objective function at each iteration of the classical optimization algorithm.

\section{Quantum Lyapunov control}
\label{Sec:QLC}

Quantum Lyapunov control (QLC) is a local-in-time method for identifying controls to asymptotically steer the dynamics of a quantum system towards a desired objective \cite{PhysRevLett.69.2172, doi:10.1063/1.467132, SUGAWARA1995113, OHTSUKI1998627, A902103E, doi:10.1063/1.1559680,1272601, Mirrahimi2005ReferenceTT, doi:https://doi.org/10.1002/9780470431917.ch2}. The controls are identified utilizing a feedback law, which is derived from a suitable control Lyapunov function \cite{10.5555/545735}, chosen to capture the target objective. In this section, we describe the theory of QLC and outline certain results from the literature pertaining to its asymptotic convergence behavior. We begin by considering a quantum system whose dynamics are governed by
\begin{equation}
    i\frac{d}{d t}|\psi(t)\rangle = (\Hp+\Hd\beta(t))|\psi(t)\rangle
    \label{SchrodingerEqn}
\end{equation}
where $|\psi(t)\rangle$ is the system state vector, we have set $\hbar = 1$, and $\Hp$ and $\Hd$ denote the (unitless) ``drift'' and ``control'' Hamiltonians, and the latter couples a scalar, time-dependent control function $\beta(t)$ to the system. In this article, we choose our QLC objective to be the minimization of $\langle \Hp \rangle = \langle \psi(t)| \Hp |\psi(t)\rangle$, and thus seek a QLC strategy for designing $\beta(t)$ to accomplish this. We proceed by defining a Lyapunov function 
\begin{equation}
\Ep(|\psi(t)\rangle) = \langle\psi(t)|\Hp|\psi(t)\rangle
\label{Eq:DefineLFunc}
\end{equation}
to capture our QLC objective. Then, to minimize $\Ep$ we seek to design $\beta(t)$ such that the QLC condition
\begin{equation}
\frac{d}{dt}\Ep\leq 0, \quad \forall t \geq 0
\label{Eq:ddt}
\end{equation}
is satisfied. There is significant flexibility in choosing $\beta(t)$ to satisfy Eq.~(\ref{Eq:ddt}). Namely, given that 
\begin{equation}
\begin{aligned}
    \frac{d\Ep}{dt} &= \langle\psi(t)|i[\Hp+\beta(t)\Hd,\Hp]|\psi(t)\rangle\\
    &=\langle\psi(t)|i[\beta(t) \Hd,\Hp]|\psi(t)\rangle\\
    &=\langle\psi(t)|i[ \Hd,\Hp]|\psi(t)\rangle \beta(t)\\
    &= A(t) \beta(t)\,,
\end{aligned}
\label{eq:dEdt}
\end{equation}
where 
\begin{equation}
    A(t)\equiv \langle\psi(t)| i[\Hd, \Hp] |\psi(t)\rangle\,,
\end{equation} 
we may take
\begin{equation}
    \beta(t) = -w \, f(t,A(t))\,,
    \label{Eq:betaQLC}
\end{equation}
for $w > 0$, where $f(t,A(t))$ is any continuous function with $f(t,0)=0$ and $A(t)f(t,A(t))>0$ for all $A(t)\neq 0$ \cite{lyapunovsurvey}. This choice of $\beta(t)$ guarantees that $\Ep$ will decrease monotonically over time. When $\beta(t)$ is chosen according to Eq.~(\ref{Eq:betaQLC}), the system dynamics are governed by \begin{equation}
    i\frac{d}{d t}|\psi(t)\rangle = (\Hp+\Hd\beta(t,A(t)))|\psi(t)\rangle \, ,
    \label{Eq:NLSE}
\end{equation}
which are highly nonlinear, due to the dependence of $\beta(t)$ on the state $|\psi(t)\rangle$ via $A(t)$. 

We refer to Eq.~(\ref{Eq:betaQLC}) as a \emph{feedback law}, as it relies on feedback in order to evaluate the observable expectation value $A(t)$. Conventionally, QLC laws like Eq.~(\ref{Eq:betaQLC}) are used in simulations to design open-loop control laws; that is, they cannot be applied directly in experiments as-is, as the destructive, real-time measurements required to estimate $A(t)$ would lead to a collapse of the system state. This distinguishes QLC from real-time feedback control. 

Ideally, designing $\beta$ per Eq.~(\ref{Eq:betaQLC}) would result in asymptotic convergence to the global minimum of $\Ep$, and it has been shown that this behavior can be guaranteed when a set of sufficient conditions are met \cite{1272601, lyapunovsurvey, BEAUCHARD2007388, doi:10.1002/rnc.1748}. However, these conditions are very stringent (see Appendix \ref{AppA:Convergence}). When the protocol we construct in Sec.~\ref{Sec:FALQON} is applied to the MaxCut problem, per Sec.~\ref{Sec:AppstoMaxCut}, they are not satisfied. Nonetheless, asymptotic convergence to the global minimum can still be obtained in such settings (e.g., as illustrated in Fig.~\ref{PRAMainResults}), and in situations where convergence is not obtained, a variety of techniques can be employed improve control performance, as discussed in the following subsections.

\subsection{Inclusion of reference perturbation in $\beta(t)$}\label{Sec:RefPert}

The inclusion of a reference perturbation in $\beta(t)$ can improve the likelihood of asymptotic convergence to the global minimum of $\Ep$ \cite{BEAUCHARD2007388, doi:10.1002/rnc.1748, lyapunovsurvey}. Here, we consider the inclusion of a reference perturbation $\lambda(t)$ such that the time-dependent system Hamiltonian is given by
\begin{equation}
H(t) = \Hp+(\lambda(t)+\beta(t))\Hd\,.
\label{tdepH}
\end{equation}
 Inspecting Eq.~(\ref{tdepH}), we may define system (a) as a system with drift Hamiltonian $\Hp$, control Hamiltonian $\Hd$, and control function ($\lambda(t)+\beta(t)$). Meanwhile, we may define system (b) as a perturbed system with time-dependent drift Hamiltonian 
\begin{equation}
    \Hpb(t) \equiv \Hp + \lambda(t) \Hd\,,
\end{equation}
control Hamiltonian $\Hd$, and control function $\beta(t)$. Within (b), we may define the perturbed Lyapunov function
\begin{equation}
\Epb(|\psi(t)\rangle) = \langle\psi(t)|\Hpb(t)|\psi(t)\rangle\,,
\label{tdeplyap}
\end{equation}
and seek a control law that will ensure
\begin{equation}
    \frac{d \Epb}{dt}\leq 0\,,
    \label{ddt(b)}
\end{equation}
while at the same time, ideally improving convergence to the minimum of our original objective $\Ep$. At this stage, it is important to note that in practice, $\lambda(t)$ can be defined explicitly as a desired function of $t$, or implicitly as a function of $|\psi(t)\rangle$, $\Ep$, or $\Epb$. For practical reasons, we restrict our attention to the former case; for further details on the latter, we refer to refs.~\cite{BEAUCHARD2007388, doi:10.1002/rnc.1748}. 

When $\lambda(t)$ is defined as an explicit function of time, the left-side of Eq.~(\ref{ddt(b)}) is given, conveniently, by,
\begin{equation}
\begin{aligned}
\frac{d}{dt}\Epb &= \langle\psi(t)|i [H(t),\Hpb(t)] |\psi(t)\rangle \\
&= \langle\psi(t)|i [\Hpb(t)+\beta(t)\Hd,\Hpb(t)] |\psi(t)\rangle \\
&= \langle\psi(t)|i [\beta(t)\Hd,\Hpb(t)] |\psi(t)\rangle\\
&= \langle\psi(t)|i [\Hd,\Hp] |\psi(t)\rangle \beta(t)\\
&=A(t)\beta(t)
\end{aligned}
\end{equation}
which implies that even after the inclusion of the reference perturbation $\lambda(t)$ and the definition of a perturbed Lyapunov function $\Epb$, we may nonetheless define the control law for $\beta(t)$ as before, as $ \beta(t) = -w\,f(t,A(t))$, to ensure Eq.~(\ref{ddt(b)}) is satisfied. 

Within this framework, if system (b) converges to the ground state of $\Hpb(t)$ at a terminal time $t=T$, and $\lambda(T)=0$, then system (b) becomes system (a), such that $\Hpb(T) = \Hp$, meaning that the ground state of $\Hpb(t)$ is also the ground state of $\Hp$, and convergence to the desired state has been obtained. As such, it is often practical to select $\lambda(t)$ to be a slowly-varying function that tends to 0 as $t\rightarrow T$. 

\subsection{Iterative quantum Lyapunov control}\label{Sec:IterativePert}

Another technique to improve the likelihood of asymptotic convergence to the minimum of $\Ep$ is to use an iterative procedure for refining the QLC control function $\beta(t)$ \cite{Mirrahimi2005ReferenceTT}. We emphasize that the iterative QLC procedure outlined in this section is conceptually distinct from the iterative optimization procedure utilized in QOC, as the iterations involved do not involve updates determined by an optimization routine. Instead, $\beta(t)$ is updated using a QLC-derived control law, in the following manner. 

We begin by considering a system initialized as $|\psi(t=0)\rangle=|\psi_0\rangle$, and then design a control field $\beta^{(0)}(t)$ to control $\Ep$ using QLC, as per the control law of Eq.~(\ref{Eq:betaQLC}), over some fixed time interval $t\in[0,T]$. We denote the trajectory of $\Ep$ over this time interval by $\Ep^{(0)}(t)$, and denote the associated state by $|\psi^{(0)}(t)\rangle$. Subsequent steps are then carried out as follows. For iterations $j\geq 1$, $\beta^{(j-1)}(t)$ serves as a reference perturbation, as denoted by $\lambda(t)$ in Sec.~\ref{Sec:RefPert}. Then, $\beta^{(j-1)}(t)$, $\Ep^{(j-1)}(t)$, and $|\psi^{(j-1)}(t)\rangle$ all describe the dynamics of a perturbed system (b), whose time-dependent Hamiltonian is
\begin{equation}
    \Hpb^{(j-1)}(t) = \Hp+\beta^{(j-1)}(t)\Hd\,.
\end{equation}
A new QLC field $\tilde{\beta}^{(j)}(t)$ is then determined for $t\in[0,T]$ using the framework in Sec.~\ref{Sec:RefPert}, where a perturbed Lyapunov function based on $\Hpb^{(j-1)}$ is utilized, and $\tilde{\beta}^{(j)}(t)$ is chosen according to Eq.~(\ref{Eq:betaQLC}). After $\tilde{\beta}^{(j)}(t)$ has been computed for $t\in[0,T]$, the update rule is given by
\begin{equation}
\beta^{(j)}(t) = \beta^{(j-1)}(t)+\tilde{\beta}^{(j)}(t).
\end{equation}

For $T$ chosen to be large enough for the perturbed system to converge to the unperturbed system at each iteration, i.e., such that $\beta^{(j)}(T)=0$, causing $\Hpb^{(j)}(T) = \Hp$, $\forall j$, this procedure guarantees a monotonic improvement of $\Ep(T)$ with respect to iteration, as per,
\begin{equation}
    \begin{aligned}
    \langle\psi^{(j)}(T)|\Hp|\psi^{(j)}(T)\rangle & = \langle\psi^{(j)}(T)|\Hpb^{(j-1)}(T)|\psi^{(j)}(T)\rangle\\
    &\leq \langle\psi^{(j)}(0)|\Hpb^{(j-1)}(0)|\psi^{(j)}(0)\rangle \\
    & =\langle\psi_0|\Hpb^{(j-1)}(0)|\psi_0\rangle \\
    & =\langle\psi^{(j-1)}(0)|\Hpb^{(j-1)}(0)|\psi^{(j-1)}(0)\rangle \\
    & =\langle\psi^{(j-1)}(T)|\Hpb^{(j-1)}(T)|\psi^{(j-1)}(T)\rangle \\
    & =\langle\psi^{(j-1)}(T)|\Hp|\psi^{(j-1)}(T)\rangle ,
    \end{aligned}
    \label{Eq:IterProof}
\end{equation}
such that $\Ep^{(j)}(T)\leq \Ep^{(j-1)}(T)$ \cite{Mirrahimi2005ReferenceTT}. We note that in line 5 of the above, we have utilized the fact that $\frac{d}{dt} \langle\psi^{(j-1)}(t)|H_{p,(b)}^{(j-1)}(t)|\psi^{(j-1)}(t)\rangle = 0$, due to the fact that $|\psi^{(j-1)}(t)\rangle$ evolves under $H_{p,(b)}^{(j-1)}(t)$. 

\subsection{Extensions to multiple control functions}\label{Sec:MultiHd}

The framework of QLC can be extended in a straightforward manner to settings with multiple control functions, i.e., where the system Hamiltonian is given by
\begin{equation}
    H(t) = \Hp + \sum_j \beta(j,t) \Hdj,
\end{equation}
where $\beta(j,t)$ denotes the value of the control function that scales the $j$-th control Hamiltonian $\Hdj$ at time $t$. Then, in order to satisfy the QLC condition that $\frac{d}{dt}\Ep\leq 0$, we see that
\begin{equation}
\begin{aligned}
    \frac{d}{dt} \langle \psi(t)|\Hp|\psi(t)\rangle &= \langle\psi(t)|i[\Hp+\sum_j \beta(j,t) \Hdj, \Hp]|\psi(t)\rangle\\
    &=\langle\psi(t)|i[\sum_j \beta(j,t) \Hdj,\Hp]|\psi(t)\rangle\\
    &=\sum_j\langle\psi(t)|i[ \Hdj,\Hp]|\psi(t)\rangle \beta(j,t)\\
    &=\sum_j A(j,t) \beta(j,t)\,,
\end{aligned}
\end{equation}
where
\begin{equation}
    A(j,t)\equiv \langle\psi(t)|i[\Hdj,\Hp]|\psi(t)\rangle\,.
\end{equation}
As such, the following control laws may be used:
\begin{equation}
    \beta(j,t) = -w_jf_j(t,A(j,t)), \quad \forall j
\end{equation}
to ensure that $\Ep$ decreases monotonically over time. We remark that in cases with multiple control functions, reference perturbations may be included for any $\beta(j,t)$, following the framework outlined in Sec.~\ref{Sec:RefPert}, and iterative QLC schemes can also be used, following the framework of Sec.~\ref{Sec:IterativePert}.

\section{Feedback-based algorithm for quantum optimization}
\label{Sec:FALQON}

We now consider how the QLC framework outlined in Sec.~\ref{Sec:QLC} (eqs.~\eqref{Eq:ddt} -- \eqref{Eq:betaQLC}) can be translated into FALQON, a feedback-based algorithm for minimizing the expectation value of a Hamiltonian, that can be implemented on quantum devices. To this end, we now assume that $\Hp$ is the problem Hamiltonian that encodes a combinatorial optimization problem of interest, noting that when defined this way,
\begin{equation}
    \Ep = \langle \psi(t)|\Hp|\psi (t)\rangle
\end{equation}
may not meet all of the requirements of being a true Lyapunov function, such as positive definiteness. Now, without loss of generality, we consider alternating, rather than concurrent, applications of $\Hp$ and $\Hd$, such that the state $|\psi(t)\rangle$ undergoes a time evolution of the form 
\begin{equation}
    U = \Ud(\beta_\ell) \Up \cdots  \Ud(\beta_1) \Up\,,
    \label{eq:FALQON-sequence}
\end{equation}
where
\begin{equation}
    \Ud(\beta_k)\equiv e^{-i\beta_k \Hd\Delta t}
    \label{eq:driverunitary}
\end{equation}
and
\begin{equation}
    \Up\equiv e^{-i \Hp \Delta t}
    \label{eq:problemunitary}
\end{equation}
and $\beta_{k} = \beta((2k-1)\Delta t)$ for $k=1, 2, \cdots, \ell$, such that after each period of $\Delta t$, the Hamiltonian that is applied alternates between $\Hp$ and $\Hd$. For small $\Delta t$, this yields a Trotterized approximation to the time evolution that would be achieved in Eq.~(\ref{SchrodingerEqn}). To ensure that Eq.~(\ref{Eq:ddt}) is satisfied, we may again define $\beta$ from Eq.~(\ref{Eq:betaQLC}). In this work, we use
\begin{equation}
    w=1,\quad f(t,A(t)) = A(t)\,,
\end{equation}
such that in the alternating framework
\begin{equation}
    \beta_{k+1}= -A_k\,,
    \label{Eq:Discretebeta}
\end{equation}
where $A_k = \langle\psi_k| i[\Hd, \Hp] |\psi_k\rangle$, and $|\psi_k\rangle = |\psi(2k\Delta t)\rangle$. Importantly, we note that it is always possible to select $\Delta t$ small enough such that Eq.~(\ref{Eq:ddt}) is satisfied when the control law in Eq.~(\ref{Eq:Discretebeta}) is used (see Sec.~\ref{Selectingdt}). However, if $\Delta t$ is chosen to be too large, the condition in Eq.~(\ref{Eq:ddt}) can be violated. 

The implementation of this alternating procedure on a qubit device can be accomplished according to the steps in Algorithm \ref{algo1}. The preliminary step is to seed the procedure by setting $\beta_1 = \beta_{\mathrm{init}}$, and we use $\beta_{\mathrm{init}} = 0$. Then, a set of qubits are initialized in a fixed initial state $|\psi_0\rangle$, and a single circuit ``layer'' is implemented to prepare the state
\begin{equation}
    |\psi_1\rangle = \Ud(\beta_1)\Up|\psi_0\rangle\,.
\end{equation}
Next, the qubits are then measured in order to estimate $A_1$. This can be accomplished by expanding $A_1$ in the Pauli operator basis as 
\begin{equation}
    A_1 = \langle \psi_1|i[\Hd, \Hp]|\psi_1\rangle= \sum_{j=1}^N \alpha_j\langle\psi_1| P_j|\psi_1\rangle \, ,
    \label{Eq:Aexpansion}
\end{equation}
where $\alpha_j$ are scalar coefficients and $P_j$ are Pauli basis operators. We note that the number of Pauli basis operators $N$ in the expansion depends on the structure of $\Hp$ and $\Hd$ (see Eq.~(\ref{Eq:N}) below). Each $P_j$ can then be measured, and the measurements can be repeated to collect sufficiently many samples to estimate the associated expectation values. Finally, the resultant expectation values of each $P_j$ can then be used to evaluate the weighted sum in Eq.~(\ref{Eq:Aexpansion}) to estimate $A_1$. Following this, the result is ``fed back'' to set $\beta_2 = -A_1$ (or, more precisely, $\beta_2$ is set to be the negative of the approximation of $A_1$). 

For subsequent steps $k = 2, \cdots, \ell$, the same procedure is repeated: the qubits are initialized in the state $|\psi_0\rangle$, after which $k$ layers are applied to obtain 
\begin{equation}
    |\psi_k\rangle = \Ud(\beta_{k}) \Up\cdots \Ud(\beta_1) \Up |\psi_0\rangle\,.
    \label{Eq:Alternating}
\end{equation}
Then, the qubits are measured to estimate $A_{k}$ using the same procedure described above, and the result is fed back to set the value of $\beta_{k+1}$. By design, this procedure causes $\langle \Hp\rangle$ to decrease layer-by-layer as per 
\begin{equation}
    \langle\psi_1| \Hp |\psi_1\rangle\geq \langle\psi_2| \Hp |\psi_2\rangle \geq \cdots\geq \langle\psi_\ell| \Hp |\psi_\ell\rangle\,,
\end{equation}
such that the quality of the solution to the combinatorial optimization problem monotonically improves with circuit depth. The protocol can be terminated when the value of $\langle \Hp\rangle$ converges (i.e., stops decreasing), as determined via measurements, or when a threshold number of layers $\ell$ is reached. At that point, the set of $\beta$ values $\{\beta_k\}_{k=1}^\ell$ is recorded as the output.

After Algorithm \ref{algo1} concludes, the set $\{\beta_k\}_{k=1}^\ell$ can subsequently be used to prepare the state $|\psi_\ell\rangle$ in post-processing steps as needed, e.g., in order to estimate the value of $\langle \Hp\rangle$, by measuring $|\psi_\ell\rangle$ and repeating the experiment enough times to ensure reliable statistics. In addition, the associated $\ell$-layer quantum circuit can also be implemented in order to estimate the bit string $z = z_1 z_2 \cdots z_n$ associated with the best candidate solution to the underlying combinatorial optimization problem. This latter task can be accomplished by sampling the bit string $z_1 z_2\cdots z_n$ from the output distribution associated with the output state $|\psi_\ell\rangle$, i.e., by measuring $z_j = \langle\psi_\ell|Z_j|\psi_\ell\rangle$ for $j=1,\cdots,n$ and then concatenating the results to form
\begin{equation}
  z = z_1 z_2 \cdots z_n,
\end{equation}
where $Z_j$ denotes the Pauli operator acting on qubit $j$ \footnote{For applications of FALQON to MaxCut on regular graphs, as studied in Sec.~\ref{Sec:AppstoMaxCut}, the results of this procedure for measuring the bit string $z_1z_2\cdots z_n$ will be concentrated around the mean when shallow circuits are used. 
The proof for concentration of $\langle \Hp\rangle$ for fixed $\ell$ follows directly from the proof in Sec. III of \cite{2014arXiv1411.4028F}. Given that $\Hp$ and $Z_1Z_2\cdots Z_n$ are both diagonal in the measured, computational basis, it follows that concentration holds for the latter as well.}. After collecting a set of samples, the bit string associated with the best solution to the combinatorial optimization problem, i.e., the bit string $z$ that returns the minimum value of the associated cost function $C(z)$, should be saved as the best approximate solution to the combinatorial optimization problem of interest. 

\begin{algorithm}[H]
\caption{FALQON \label{algo1}}
\begin{algorithmic}[1]
\STATE \textbf{set} $\Hp$, $\Hd$, $\Delta t$, $\ell$, $|\psi_0\rangle$
\STATE Seed the procedure 

$\qquad \beta_1 \leftarrow 0$
\STATE Initialize the qubits 

$\qquad |\psi\rangle \leftarrow |\psi_0\rangle$ 

\STATE Implement 1 layer 

$\qquad |\psi_1\rangle \leftarrow \Ud(\beta_1) \Up|\psi_0\rangle$
\STATE Estimate the value of $A_1$ by measuring the qubits in the state $|\psi_1\rangle$ and repeating the experiment enough times to ensure reliable statistics. 
\STATE $\beta_2\leftarrow - A_1$ 
\STATE $k \leftarrow 1$
\WHILE{$k<\ell$}
\STATE $k \leftarrow k+1$
\STATE Initialize the qubits 

$\qquad |\psi\rangle \leftarrow |\psi_0\rangle$ 
\STATE Implement $k$ layers 

$\qquad |\psi_{k} \rangle\leftarrow \Ud(\beta_{k})\Up \cdots \Ud(\beta_1)\Up|\psi_0\rangle$
\STATE Estimate the value of $A_{k}$ by measuring the qubits in the state $|\psi_k\rangle$ and repeating the experiment enough times to ensure reliable statistics. 
\STATE $\beta_{k+1}\leftarrow -A_k$ 
\ENDWHILE
\STATE \textbf{output} $\{\beta_k\}_{k=1}^\ell$
\end{algorithmic}
\end{algorithm}

We note that FALQON has similarities to other quantum circuit parameter-setting protocols that involve ``greedy'', layer-by-layer optimization, e.g., where a classical optimization routine is used to sequentially optimize quantum circuit parameters in order to minimize a cost function in a layer-wise manner \cite{carolan2020variational, skolik2021layerwise, PhysRevA.103.032607,campos2021training}. In fact, the parameter-setting rule given in Eq.~(\ref{Eq:Discretebeta}) corresponds to taking a step ``down'' in the direction of the local gradient $\frac{d}{d\beta_k}\Ep$ with a step size of $\Delta t$, thus suggesting that there is a natural connection between FALQON and layer-wise circuit optimization methods that proceed by gradient descent \cite{verdon2019quantum}. We also remark that the ADAPT-QAOA approach developed in ref.~\cite{2020arXiv200510258Z} has certain similarities to FALQON, e.g., it also utilizes information about $\frac{d}{dt} \langle \psi(t)| \Hp |\psi(t)\rangle$ to step forward from layer to layer. However, their stepping procedure involves selecting from a set of driver Hamiltonians, while still containing a classical optimization loop.

Having outlined FALQON, we now turn to the prospect of boosting its performance using the techniques outlined in Secs. (\ref{Sec:RefPert}), (\ref{Sec:IterativePert}), and (\ref{Sec:MultiHd}). We begin by discussing how a reference perturbation may be introduced, as per Eq.~(\ref{tdepH}) and how the framework outlined in Sec. (\ref{Sec:RefPert}) may be adapted to the quantum device setting. As before, this can be accomplished by simply ``Trotterizing'' Eq.~(\ref{tdepH}), and implementing a quantum circuit with the form
\begin{equation}
\Ud(\nu_\ell)\Up\cdots \Ud(\nu_1)\Up\Ud(\nu_0)\Up
\end{equation}
where $\nu_k = \lambda_k+\beta_k$, where $\lambda_k$ is the value of the reference perturbation at the $k$-th layer and $\beta_k$ is the value of the control parameter at the $k$-th layer, determined via $\beta_k = A_{k-1}$. Numerical illustrations showing how the performance of FALQON can be improved with the inclusion of a reference perturbation can be found in \cite{magann2021feedbackbased}. In a similar fashion, the iterative QLC procedure discussed in Sec.~\ref{Sec:IterativePert} can also be adapted the context of quantum optimization, in order to successively improve the quality of the solutions obtained. After applying a first-order Trotter decomposition, the circuits at the $j$-th iteration will have the structure 
\begin{equation}
\Ud(\beta_\ell^{(j)})\Up\cdots \Ud(\beta_1^{(j)})\Up\Ud(\beta_0^{(j)})\Up \, ,
\end{equation}
where $\beta_k^{(j)} = \beta_k^{(j-1)}(t)+\tilde{\beta}_k^{(j)}(t)$ for $k = 0,\cdots,\ell$, and the iterative procedure is seeded by determining $\beta^{(0)}$ via Algorithm \ref{algo1}. Finally, the approach discussed in Sec. (\ref{Sec:MultiHd}) may also be extended to the quantum device setting, by modifying the layered quantum circuit structure to include evolutions under additional driver Hamiltonians. 

\subsection{Selecting $\Delta t$ \label{Selectingdt}} 

We now consider the selection of the time step $\Delta t$ in order to ensure that Eq.~(\ref{Eq:ddt}) will hold. To this end, we consider a single layer of FALQON, such that
\begin{equation}
\begin{aligned}
\Eptt &= \langle\psi(t+2\Delta t)|\Hp|\psi(t+2\Delta t)\rangle\\
&=\langle \psi(t)|e^{i\Hp\Delta t }e^{i\Hd\beta_t\Delta t}\Hp e^{-i\Hd\beta_t\Delta t}e^{-i\Hp\Delta t}|\psi(t)\rangle \, .
\end{aligned}
\end{equation}
For the following, we adopt the notation $\langle\cdot\rangle_t\equiv\langle\psi(t)|\cdot|\psi(t)\rangle$. We express each of the exponentials above using a Taylor series expansion:

\begin{equation}
\begin{aligned}
\Eptt &=\Ept+i\langle[\Hd,\Hp]\rangle_t\beta_t\Delta t-\langle[\Hp,[\Hd,\Hp]]\rangle_t\beta_t\Delta t^2\\
&\qquad-\langle[\Hd,[\Hd,\Hp]]\rangle_t\beta_t^2\Delta t^2+O(\Delta t^3)\\
&=\Eptt^{(0)}+\Eptt^{(1)}+\Eptt^{(2)}+\cdots
\end{aligned}
\end{equation}
where the superscripts label the orders of $\Delta t$. Since we only use $\beta$ to design the first order term, we would like this to be the dominant term in the expansion such that 
\begin{equation}
|\Eptt^{(1)}|> | \sum_{k=2}^\infty \Eptt^{(k)}|\, .
\label{eq:condition}
\end{equation}
This way, designing $\Eptt^{(1)}$ appropriately will enforce that $\Eptt$ decreases. The left-side of Eq.~(\ref{eq:condition}) is given by
\begin{equation}
\begin{aligned}
|\Eptt^{(1)}| &=|\langle  [\Hd,\Hp]\rangle_t|\,|\beta_t|\,\Delta t\\
&=|A_t|\,|\beta_t|\Delta t\,.
\end{aligned}
\end{equation}
Meanwhile, the magnitude of higher-order terms such as $\Eptt^{(2)}$ can be bounded as,
\begin{equation}
\begin{aligned}
|\Eptt^{(2)}&| = |\langle[\Hp,[\Hd,\Hp]]\rangle_t\beta_t+\langle[\Hd,[\Hd,\Hp]]\rangle_t\beta_t^2|\Delta t^2\\
&\leq \big(|\langle[\Hp,[\Hd,\Hp]]\rangle_t|\\
&\qquad+|\langle[\Hd,[\Hd,\Hp]]\rangle_t||\beta_t|\big)|\beta_t|\Delta t^2\\
&\leq \big(\norm{[\Hp,[\Hd,\Hp]]}+\norm{[\Hd,[\Hd,\Hp]]}|\beta_t|\big)|\beta_t|\Delta t^2\\
&\leq  \big(2\norm{\Hp\Hd\Hp}+\norm{\Hp\Hp\Hd}+\norm{\Hd\Hp\Hp}\\
&\,\, +\norm{\Hd\Hd\Hp}+\norm{\Hp\Hd\Hd}\\
&\,\,+2\norm{\Hd\Hp\Hd}|\beta_t|\big)|\beta_t|\Delta t^2\\
&\leq \big(2\norm{\Hp}^2\norm{\Hd}+2\norm{\Hd}^2\norm{\Hp}|\beta_t|\big)2|\beta_t|\Delta t^2 \\
&=2\np\nd|\beta_t| \big(2\np+2\nd|\beta_t|\big)\Delta t^2 \, ,
\end{aligned}
\end{equation}
where in the last line we introduce the abbreviated notation $\nd \equiv \norm{\Hd}$ and $\np\equiv \norm{\Hp}$. Expressions for the magnitude of any higher-order (i.e., $k \geq 2$) term can be found following the same procedure, which results in the following general expression at $k$-th order:
\begin{equation}
    |\Eptt^{(k)}|=2\np\nd|\beta_t| \big(2\np+2\nd|\beta_t|\big)^{k-1}\Delta t^k \, .
    \label{Eq:kthorder}
\end{equation}

Given Eq.~(\ref{Eq:kthorder}), the right side of Eq.~(\ref{eq:condition}) can be bounded by
\begin{equation}
\begin{aligned}
| \sum_{k=2}^\infty \Eptt^{(k)}| &\leq  \sum_{k=2}^\infty |\Eptt^{(k)}|\\
&\leq  2\np\nd|\beta_t|\sum_{k=2}^\infty (2\np+2\nd|\beta_t|)^{k-1}\Delta t^k\\
&=  \frac{\np\nd|\beta_t|}{\np+\nd|\beta_t|}\bigg(\sum_{k=0}^\infty (2\Delta t(\np+\nd|\beta_t|))^k\\
&\quad-  1-2\Delta t(\np+\nd|\beta_t|)\bigg)\,.\\
\label{Eq:wgeomseries}
\end{aligned}
\end{equation}
For $2\Delta t(\np+\nd|\beta_t|)<1$ the geometric series converges. Under this assumption, we can rewrite the condition in Eq.~(\ref{eq:condition}) as
\begin{equation}
\begin{aligned}
|A_t||\beta_t|\,\Delta t&>\frac{\np\nd|\beta_t|}{\np+\nd|\beta_t|}\bigg(\frac{1}{1-2\Delta t(\np+\nd|\beta_t|)}\\
&\quad-  1-2\Delta t(\np+\nd|\beta_t|)\bigg)\,. \\
\end{aligned}
\label{Eq:rewrittencond}
\end{equation} 
We rearrange this equation to obtain a bound for $\Delta t$:
\begin{equation}
|\Delta t| < \frac{|A_t|}{2(2\nd\np+|A_t|)(\np+\nd|\beta_t|)} \, .
\label{eq:deltat}
\end{equation}
For these values of $\Delta t$, we can confirm that the geometric series in Eq.~(\ref{Eq:wgeomseries}) does converge, because 
\begin{equation}
\begin{aligned}
1&>2\Delta t(\np+\nd|\beta_t|)\\
&=\frac{2|A_t|(\np+\nd|\beta_t|) }{2(2\nd\np+|A_t|)(\np+\nd|\beta_t|)}\\
&=\frac{|A_t| }{2\nd\np+|A_t|}
\end{aligned}
\end{equation} 
is always satisfied. Therefore, if $\Delta t$ is selected according to Eq.~(\ref{eq:deltat}), it is ensured that the QLC condition in Eq.~(\ref{Eq:ddt}) will hold, and thus, that $\Ep$ will decrease monotonically as a function of layer as desired. In practice we find that $\Delta t$ can be chosen to be much larger than the value in Eq.~(\ref{eq:deltat}) due to the looseness of the bound.

\section{Applications to MaxCut}
\label{Sec:AppstoMaxCut}

We now consider applications of FALQON towards the combinatorial optimization problem MaxCut, which aims to identify a graph partition that maximizes the number of edges that are cut. For a graph $\mathcal{G}$, with $n$ nodes and edge set $\mathcal{E}$, the MaxCut problem Hamiltonian is defined on $n$ qubits as
\begin{equation}
\Hp = -\sum_{i,j \in \mathcal{E}} \frac{1}{2} \big(1-w_{ij}Z_i Z_j\big)\,,
\label{eq:HpMaxCut}
\end{equation}
where $w_{ij}$ denote the edge weights, and for unweighted graphs, $w_{ij} = 1$ for all $i,j\in\mathcal{E}$. In our analyses involving weighted graphs, we consider random edge weights $w_{ij}$ drawn from a uniform distribution between 0 and 2, such that the average edge weight is $\overline{w}=1$, matching the unweighted case. Furthermore, we consider $\Hd$ to have the standard form 
\begin{equation}
    \Hd=\sum_{j=1}^nX_j\,.
    \label{eq:sigmaxdriver}
\end{equation}
Given these choices for $\Hp$ and $\Hd$, 
\begin{equation}
\begin{aligned}
    A_k &= \langle\psi_k|i[\Hd,\Hp]|\psi_k\rangle \\
    &= \sum_{i,j \in \mathcal{E}} w_{ij}\big(\langle \psi_k|Y_i Z_j|\psi_k\rangle +\langle\psi_k|Z_i Y_j|\psi_k\rangle\big)\,,
\end{aligned}
\label{Eq:MaxCutAExpansion}
\end{equation}
where $X_j$ and $Y_j$ denote the Pauli operators acting on qubit $j$. As such, evaluating the feedback law $\beta_{k+1}=-A_k = -\langle\psi_k|i[\Hd,\Hp]|\psi_k\rangle$ requires measuring the expectation values of 
\begin{equation}
    N\leq n(n-1)
    \label{Eq:N}
\end{equation}
Pauli basis operators (i.e., in this case, two-qubit Pauli strings), where the exact value of $N$ depends on the structure of the graph under consideration. 

In order to assess the performance of FALQON towards MaxCut, we consider two figures of merit: the approximation ratio, 
\begin{equation}
    r_{\textrm{A}} = \frac{\langle \Hp \rangle} {\langle \Hp \rangle_{\min}}\,,
\end{equation}
which is proportional to the original Lyapunov function $\Ep$, and the success probability of measuring the (potentially) degenerate ground state, 
\begin{equation}
    \phi = \sum_i|\langle \psi| q_{0,i} \rangle|^2\,,
\end{equation}
which gives the probability of obtaining the global minimum solution to the original combinatorial optimization problem. Each of these two figures of merit can take on values between 0 and 1, where $r_A=\phi=1$ corresponds to the optimal (i.e., ground state) solution.

\subsection{Numerical illustrations on 3-regular graphs}

\begin{figure}[t]
\includegraphics[width=1.0\columnwidth]{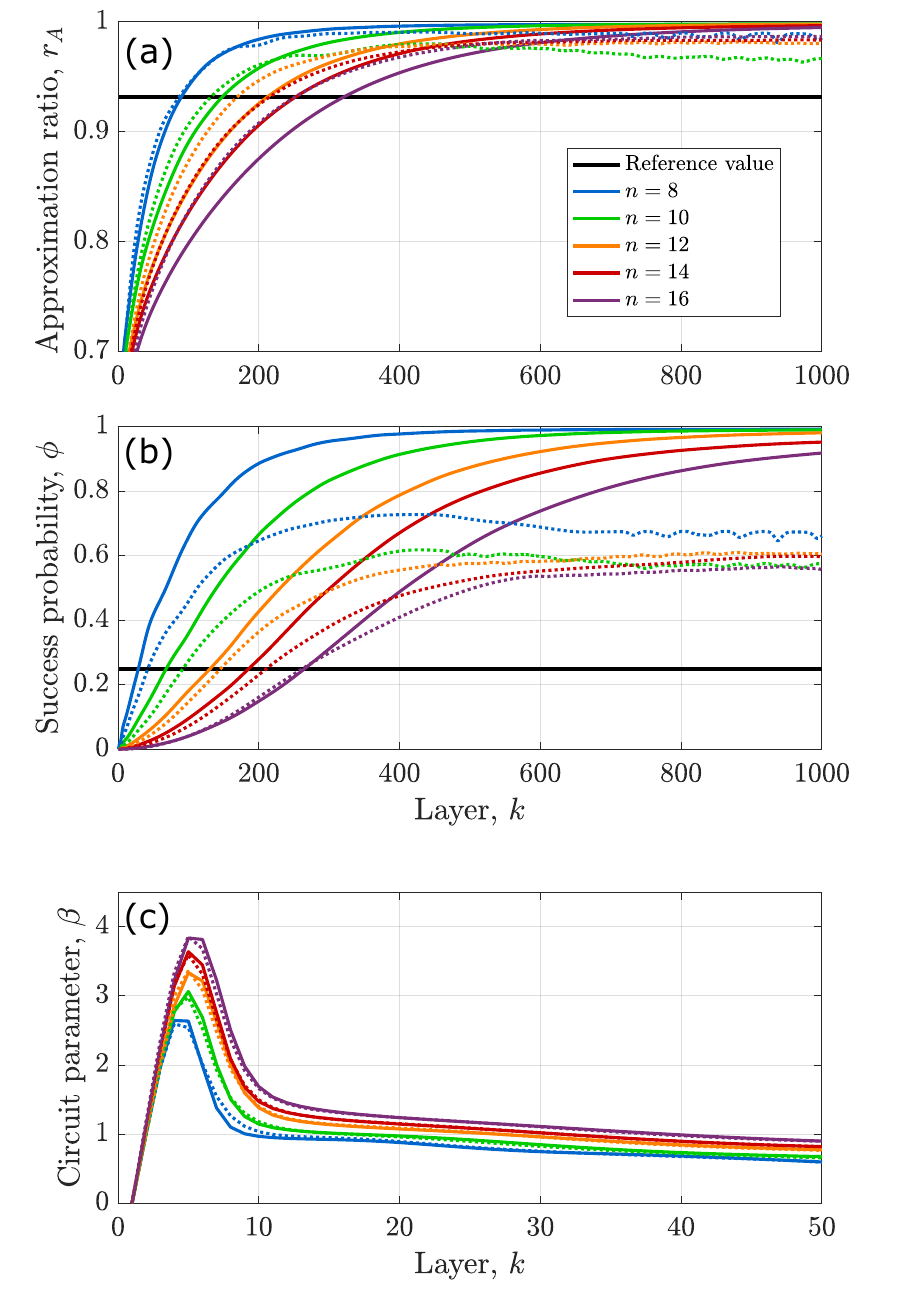}
\caption{The performance of FALQON, as quantified by the approximation ratio $r_{\textrm{A}}$ in (a) and the success probability $\phi$ in (b) is shown for different values of $n$ as a function of layer $k$, where each layer is formed by an application of the problem Hamiltonian, $\Hp$, and a driver Hamiltonian, $\Hd$, defined according to Eqs. (\ref{eq:HpMaxCut}) and (\ref{eq:sigmaxdriver}), respectively. This leads to full circuits with an alternating structure, as per Eq. (\ref{Eq:Alternating}). The solid curves of different colors show the mean results for MaxCut on unweighted 3-regular graphs over a set of different graphs at each problem size $n$, where the latter denotes the number of vertices (equivalently, the number of qubits) of the graph. For $n=8,10$, all possible graphs are considered; for $n = 12,14,16$, 50 graphs are considered at each problem size. The dotted curves show mean results for weighted 3-regular graphs, where edge weights are assigned to each graph, selected from a uniform distribution between 0 and 2. The solid black lines show the reference values $r_A = 0.932$ and $\phi = 0.25$. In (c), the mean values of $\beta$ are plotted as a function of layer.  }
\label{PRAMainResults}
\end{figure}

We now examine the performance of FALQON towards MaxCut on 3-regular graphs via a series of numerical illustrations (for a demonstration in quantum hardware, see our companion article \cite{magann2021feedbackbased}). We consider both weighted and unweighted 3-regular graphs with $n \in \{8, 10, \cdots, 20\}$ vertices. For weighted graphs, the edge weights are randomly sampled from a uniform distribution over $(0,2)$. For graphs with $n=8,10$ vertices, we consider the set of all nonisomorphic, connected 3-regular graphs. For each value of $n=12,\cdots,20$, we consider a set of 50 randomly-generated nonisomorphic graphs. The qubits are initialized in the ground state of $\Hd$, and the performance of FALQON is quantified using the mean (over the set of graphs) of $r_{\textrm{A}}$, and $\phi$. We relate the performance to two reference values: $r_{\textrm{A}} = 0.932$, corresponding to the highest approximation ratio that can currently be guaranteed using a classical approximation algorithm for MaxCut on unweighted, 3-regular graphs (i.e., the algorithm of Goemans and Williamson \cite{goemans1995improved}), and $\phi = 0.25$, which implies that on average, $4$ circuit repetitions will be needed in order to obtain a sample bit string corresponding to the ground state. The results are collected in Fig.~\ref{PRAMainResults}. 

In Fig.~\ref{PRAMainResults}(a) and (b), the $r_{\textrm{A}}$ and $\phi$ results are shown for graphs with $n=8,10,\cdots,16$ vertices, and the associated reference values are plotted in black. The solid curves show the results for unweighted graphs, while the dotted curves of corresponding color show the results for weighted graphs. We find that FALQON has superior performance on unweighted graphs, given that $r_A$ and $\phi$ appear to converge to higher values. Nonetheless, FALQON does lead to monotonic convergence towards high $r_{\textrm{A}}$ values as a function of layer for weighted graphs as well. However, for weighted graphs, we find many instances where $\phi$ fails to converge to 1, as shown in Fig.~\ref{PRAMainResults}(b). Like behavior has been found in numerical studies of QAOA, where the inclusion of edge weights in MaxCut leads to the appearance of additional poor-quality local minima in the optimization landscape \cite{shaydulin2022parameter}. Meanwhile, in Fig.~\ref{PRAMainResults}(c), we plot the associated values of $\beta$ up to layer $k=50$, where the solid and dotted curves correspond to the results for unweighted and weighted 3-regular graphs, respectively. We find that there is strong agreement between the $\beta$ curves for unweighted and weighted graphs at each problem size $n$. Furthermore, all $\beta$ curves exhibit a very consistent shape. 

In these illustrations, the only free parameter is the time step $\Delta t$. This is tuned to be as large as possible for each value of $n$, a value we call the critical time step and denote by $\Delta t_c$, as long as the condition in Eq.~(\ref{Eq:ddt}) is met for all (unweighted) problem instances considered up to 1,000 layers. We then utilize the same value of $\Delta t$ for the weighted graphs at each problem size, noting that in some weighted instances, this leads to a violation of Eq.~(\ref{Eq:ddt}), and subsequent non-monotonic behavior of $r_A$ and $\phi$. For a closer look at how violations of Eq.~(\ref{Eq:ddt}) can manifest in individual problem instances, we refer to Appendix \ref{AppA}).

\begin{figure}[t]
\includegraphics[width=1.0\columnwidth]{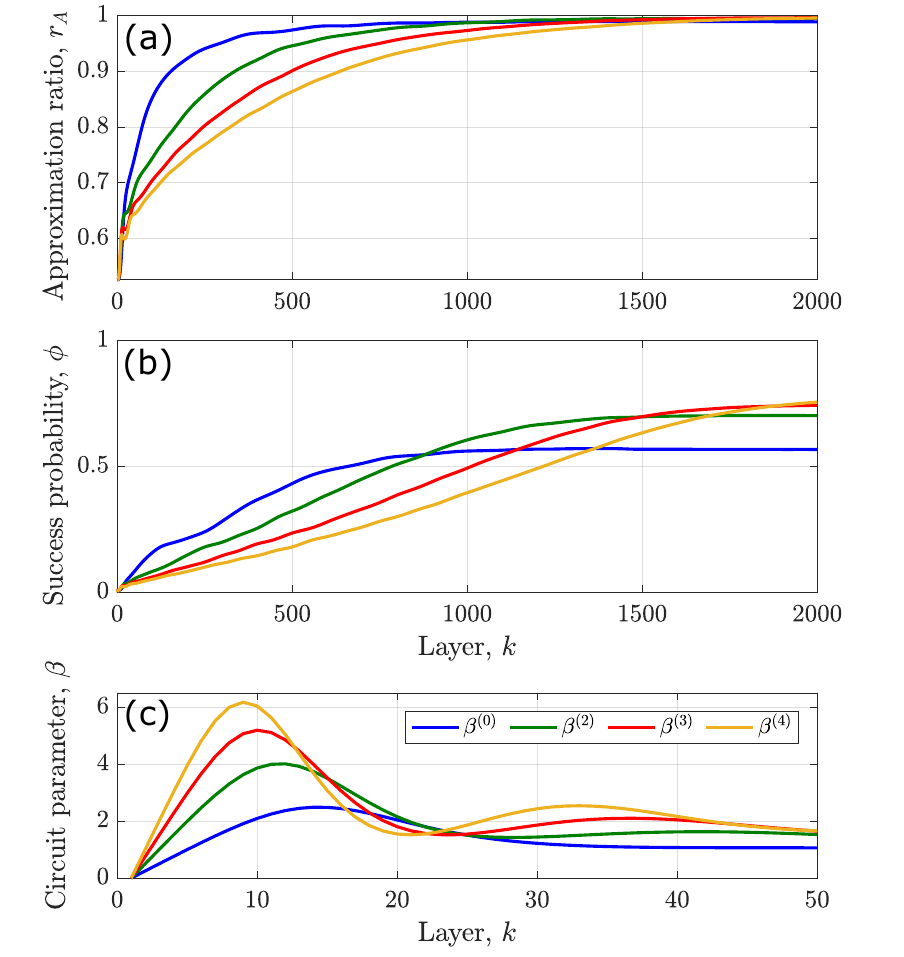}
\caption{Performance of FALQON using the iterative QLC approach, applied to a weighted 3-regular graph with $n = 8$ vertices. Three iterations are shown, using a time step of $\Delta t = 0.012$. In (a) and (b), the approximation ratio, $r_{\textrm{A}}$ and the success probability, $\phi$ are plotted as a function of layer, respectively. Panel (c) shows how the associated $\beta$ curves are refined at each iteration of the procedure. }
\label{IterativePert}
\end{figure}

Noting that convergence can be more challenging for weighted instances of MaxCut, we next explore how the performance of FALQON can be improved when it is modified using the iterative QLC heuristic introduced in Sec.~\ref{Sec:IterativePert}, with results presented in Fig.~\ref{IterativePert}. We apply this iterative QLC heuristic to a weighted instance of a 3-regular graph with $n=8$ vertices, where the base FALQON algorithm displays good convergence with respect to $r_A$, but where $\phi$ fails to reach high values, and asymptotes to only $\phi\approx 0.57$. In Fig.~\ref{IterativePert}(c), we show the $\beta$ curves that result from three iterations of the procedure, while panel (b) shows how these iterations serve to improve the convergence of $\phi$. We refer to Ref.~\cite{magann2021feedbackbased} for an illustration of the improvement provided by the reference perturbation heuristic and also another random perturbation heuristic motivated by simulated annealing.

\subsection{Behavior under measurement noise}

\begin{figure}
\includegraphics[width=1.0\columnwidth]{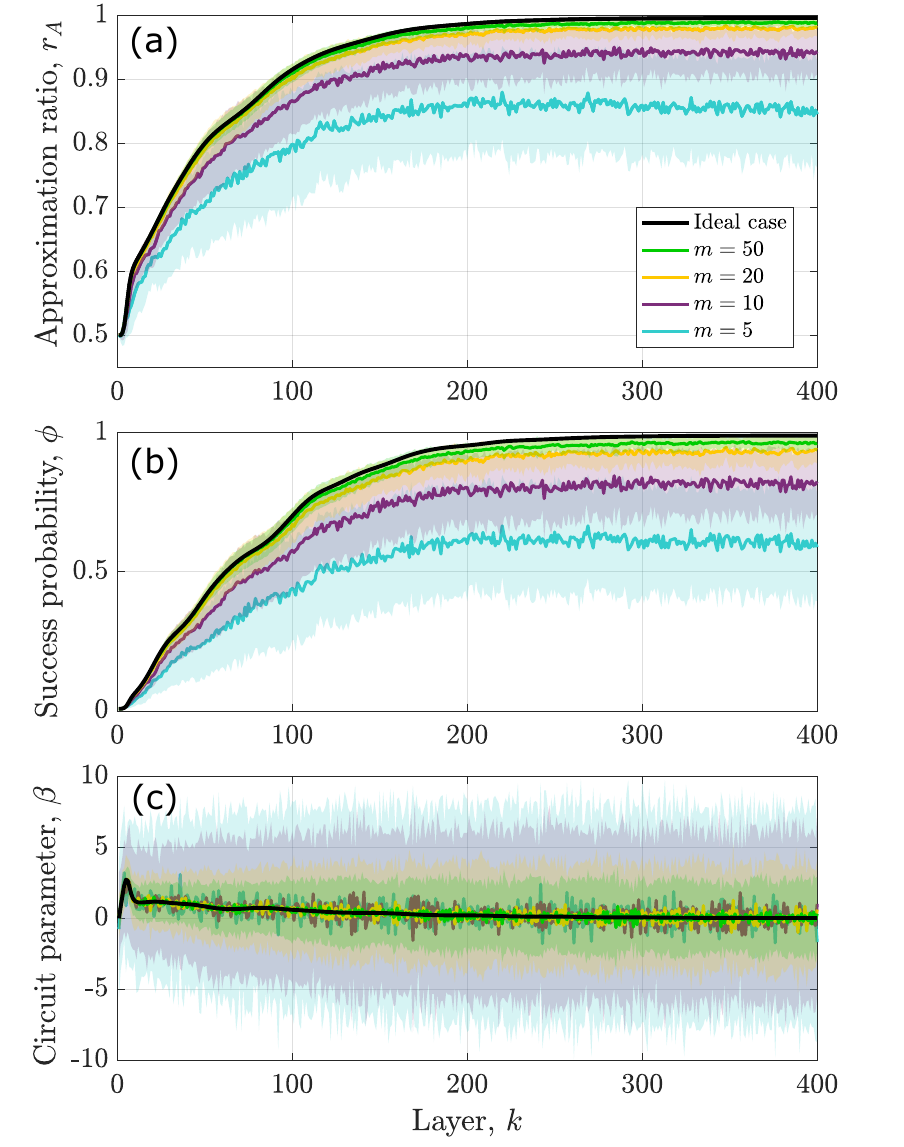}
\caption{Performance of FALQON on an instance of MaxCut on an unweighted, 3-regular graph with $n=8$ vertices in the presence of sampling noise. The time step is set to $\Delta t = 0.034$. In panel (a), the approximation ratio $r_{\textrm{A}}$ is plotted as a function of layer when $m = 2,5,20$, and 50 measurement samples are used to evaluate each of the expectation values $A_k$, $k=1,\cdots,400$. The solid black curves show the ideal, noiseless reference case. The remaining solid curves represent the mean taken over 100 realizations of this sampling process using different numbers of measurement samples, $m$; the shading shows the associated standard deviation over these realizations. Panels (b) and (c) show analogous results for the success probability, $\phi$, of measuring the 2-degenerate ground state, and the circuit parameter, $\beta$, respectively.  }
\label{robustness}
\end{figure}

Here, we analyze the performance of FALQON under measurement noise, which affects each $A_k$ value, and consequently, each $\beta_k$ value as well. This type of noise enters due to the fact that in practice, a finite number of measurement samples $m$ are used to estimate each $A_k$. We simulate this effect by sampling measurement outcomes from a multinomial distribution, defined at layer $k$ as the probability distribution over the set of eigenvalues of $i[\Hd,\Hp]$ when in the state $|\psi_k\rangle$ \footnote{We remark that in practice, it may not be possible to sample from the set of eigenvalues of $i[\Hd,\Hp]$ by measuring the observable $i[\Hd,\Hp]$ directly. Instead, the observable $i[\Hd,\Hp]$ may be expanded as a weighted sum of terms as per Eqs. (\ref{Eq:Aexpansion}) and (\ref{Eq:MaxCutAExpansion}). Then, the terms can then be grouped into sets of observables that can be measured jointly. }. 
The results are collected in Fig.~\ref{robustness}, which shows the performance of FALQON when $m=2,5,20$ and 50 samples are used to estimate $A_k$, for an instance of MaxCut on a 3-regular graph with 8 vertices. The results shown are representative of the behavior across other instances we studied. Our findings suggest that FALQON is robust to the effects of sampling noise, and can be effective even in the presence of significant measurement noise. We also find that as the number of samples $m$ that are used decreases, performance improves if $\Delta t$ is selected to decrease as well.

\subsection{Comparison with QAOA}

\begin{figure}
\includegraphics[width=1.0\columnwidth]{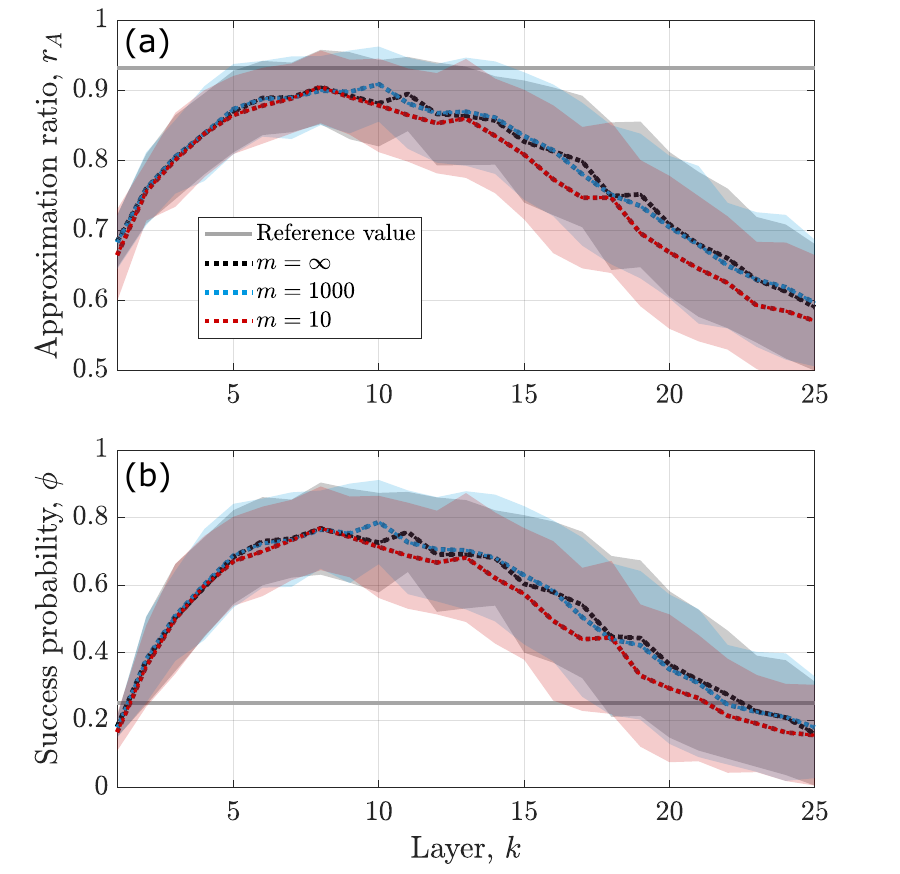}
\caption{Numerical analysis of the performance of QAOA for an instance of MaxCut on an unweighted, 3-regular graph with $n=8$ vertices; the instance of MaxCut considered here is the same that is considered in the analysis presented in Fig.~\ref{robustness}. For this analysis, we simulate implementations of QAOA that scan over $k = 1,2,\cdots,25$ layers. A series of 100 realizations are performed at each value of $k$, in which the initial circuit parameter values are selected at random, and $q=1000$ iterations of SPSA is are performed in order to optimize the circuit parameters. SPSA makes two calls to the objective function per iteration, and for each of these calls, we consider estimating the value of $\langle \Hp\rangle$ using $m=10$, $m=1000$, and $m=\infty$ samples, where the latter corresponds to the case of ideal measurements and perfect resolution of the expectation value $\langle \Hp\rangle$. The performance of QAOA for each pair of $(m,k)$ values is then quantified by the approximation ratio, $r_A$, and the success probability, $\phi$, in panels (a) and (b), respectively. In each panel, the dotted curves and shaded regions show the mean and standard deviations computed over the realizations, respectively, as a function of $k$ for each value of $m$. Meanwhile, the gray horizontal lines denote the references values of $r_A = 0.932$ and $\phi = 0.25$.}
\label{QAOAcomp}
\end{figure}

In this section, we consider how the performance of FALQON can be expected to compare with that of QAOA. We recall that a key feature of FALQON is that it does not require any classical optimization, and as a result, the resources required for FALQON are substantially different, compared with the resources required for QAOA. In particular, in Fig.~\ref{PRAMainResults} we found that when FALQON is applied to MaxCut on regular graphs, it is able to achieve high values of the approximation ratio, $r_A$, and relatively high values of the success probability, $\phi$, with no classical optimization. Furthermore, it can also achieve high values of $r_A$ and $\phi$ in the presence of measurement noise, i.e., when only a small number of measurement samples, $m$, are used to estimate the expectation values $A_k$ at each layer, as shown in Fig.~\ref{robustness}. However, it is evident from these figures that FALQON does require relatively deep circuits in order to achieve this good performance. 

We now turn to QAOA. Applications of QAOA as a hybrid quantum-classical algorithm have mostly focused on shallow circuits. In this regime, QAOA can be expected to find better solutions than FALQON, through the aid of classical optimization. The premise is that the classical optimization will allow for extracting the best attainable solution from the quantum processor within a limited circuit depth. Beyond shallow circuits, in principle QAOA is capable of achieving solutions that improve monotonically with respect to the depth of the circuit. However in practice, seeing a monotonic improvement in solution quality as the number of QAOA layers $k\rightarrow\infty$ would require resources that are not practically feasible (i.e., the ability to identify globally optimal solutions at each value of $k$ through classical optimization). In practice, scaling up QAOA to larger problem sizes and deeper circuits will cause the classical optimization cost to rise, due to the fact that optimization is harder in higher dimensions, and in fact, formally scales exponentially with increasing number of layers \cite{Bittel_Kliesch_2021}.

Because QAOA and FALQON require different resources it is difficult to compare them directly, especially with one figure of merit. Instead, in the following we analyze the performance of each separately on the same problem instance. Fig.~\ref{QAOAcomp} presents the performance of QAOA on the same instance of MaxCut as is considered in the FALQON analysis presented in Fig.~\ref{robustness}. In particular, Fig.~\ref{QAOAcomp} presents the results of QAOA simulations for circuits with $k = 1,2,\cdots,25$ layers. A series of 100 QAOA realizations are performed at each value of $k$. For each realization, the initial parameter values are chosen uniformly randomly from $\beta_k\in [0,\pi)$, $\gamma_k\in [0,2\pi)$. Then, the parameter optimization is performed using 1,000 iterations of the Simultaneous Perturbation Stochastic Approximation (SPSA) algorithm \cite{spall1998overview, spall1998implementation}, which involves perturbing $\langle \Hp\rangle$ in order to estimate an approximate gradient at each optimization iteration, and utilizing this to perform gradient descent. The gradient is approximated at each iteration by evaluating the objective function $\langle \Hp\rangle$ twice, regardless of how many optimization parameters are involved. For each of these evaluations, we consider estimating the value of $\langle \Hp\rangle$ using $m=10$, $m=1000$, and $m=\infty$ samples, where the latter corresponds to the case of ideal measurements and perfect resolution of the expectation value $\langle \Hp\rangle$. The performance of QAOA is then quantified by the approximation ratio, $r_A$, and the success probability, $\phi$, with results plotted in Fig.~\ref{QAOAcomp}(a) and (b), respectively. 

We now contrast the results in Fig.~\ref{QAOAcomp} with the earlier FALQON results in Fig.~\ref{robustness}. Recalling first the FALQON results, we find monotonic improvements in $r_A$ and $\phi$ with respect to layer, $k$. We also find that increasing $m$ leads to faster convergence. Turning to QAOA, we find that the QAOA results for different values of $m$ are not significantly different, indicating that SPSA performs comparably in the presence of different levels of sampling noise for this problem instance. We also find that the behavior of $r_A$ and $\phi$ with respect to $k$ is non-monotonic. That is, both of these figures of merit first increase as a function of $k$. This is likely due to the increased expressivity of circuits as more layers are added, i.e., better solutions become reachable, and 1,000 SPSA iterations is sufficient for exploring the parameter space and identifying these solutions. As $k$ increases further, the advantages of this increasing expressivity are subsequently counterbalanced by the fact that adding layers also adds more optimization variables, increasing the dimension of the optimization space and the difficulty of the optimization problem. The behavior of $r_A $ and $\phi$ then rolls over and begins to deteriorate as a consequence of the increasing difficulty of the optimization problem, once the limited number of optimization iterations allowed becomes insufficient for exploring the space and finding good solutions. 

We emphasize that these findings are not specific to the MaxCut problem instance analyzed here, and hold generically. Also, the choice of optimization algorithm does not significantly affect the conclusions, the same overall behavior is seen with other optimization algorithms.

These findings support the notion that QAOA is likely favorable in settings where classical optimization resources are sufficiently available and quantum resources are limited to the regime of shallow circuits. On the other hand, FALQON demonstrates strong performance for deep circuits, and does not require classical optimization resources, meaning that it has the advantage of not incurring a rising classical cost as the circuit depth is scaled up. This suggests that in cases where it is feasible to implement deep circuits, FALQON could offer a considerable advantage. 

We conclude this section with another comparison of the resources required by FALQON and QAOA, now in the context of their sampling complexity for MaxCut, as quantified by the total number of samples (i.e., circuit repetitions) that are required, denoted $N_{s}$. We denote by $m$ the number of samples needed to estimate the expectation value of a two-qubit Pauli string $P_j$, and for simplicity, $m$ is assumed to be independent of $P_j$. We first consider QAOA: given that all terms in $H_{\textrm{p}}$ commute, for $q(\ell)$ classical optimization iterations of QAOA, $N_{s} = mq(\ell)$, assuming one evaluation of $\langle H_p\rangle$ per optimization iteration. We note that for any reasonable convergence, the number of optimization iterations, $q(\ell)$ depends at least linearly on $\ell$.  However, if a gradient algorithm is used for QAOA, additional samples will be needed. We assume that for $\ell$ layers of QAOA, $2\ell$ gradient elements are required, for each of the $2\ell$ circuit parameters. Assuming that at least $m$ samples are needed to estimate each gradient element (i.e., to evaluate $\langle H_{\textrm{p}}\rangle$ for at least one perturbation of each circuit parameter), then for $q$ iterations,
\begin{equation}
N_{s}^{QAOA} \geq mq(\ell)(1+2\ell) = \mathcal{O}(mq(\ell)\ell).
\label{qaoasamp}
\end{equation}

In FALQON, additional measurements are needed to evaluate $A_1, \cdots, A_\ell$. This involves measuring each of the terms in $i[H_{\textrm{d}}, H_{\textrm{p}}]$, which contains a set of $YZ$ terms and a set of $ZY$ terms. In principle, these two sets can be combined to form a single set, given that each $Y_j Z_k$ term commutes with each $Z_j Y_k$ term and can thus be measured together \cite{2019arXiv190806942C, 2019arXiv190709386Y, doi:10.1063/1.5141458}; however, we consider the scenario that commuting $Y_i Z_j$ and $Z_i Y_j$ terms are measured separately, as per current convention, although terms such as $Y_i Z_j$ and $Z_l Y_m$, which act nontrivially on disjoint pairs of qubits, may be measured simultaneously. Then, for a graph $\mathcal{G}$ with maximum degree $d$, the expectation value of $i[H_{\textrm{d}}, H_{\textrm{p}}]$ can be estimated in maximally $2m(d+1)$ repetitions 
\footnote{We can see this by mapping the problem of estimating the number of repetitions to the problem of edge coloring, which asks for an assignment of colors to the edges of a graph such that no edges sharing a common vertex have the same color. The number of colors corresponds to the number of repetitions, as the same rule applies (i.e., that no non-commuting 2-qubit terms sharing a common qubit can be measured simultaneously). From Vizing's theorem, at most $d+1$ colors are needed for edge coloring a (simple) graph whose maximum degree is $d$ \cite{10020875344}. Although the task of determining the optimal edge coloring is in general NP-hard, classical, polynomial-time algorithms exist for assigning at most $d+1$ colors to graphs with maximum degree $d$ \cite{Misra92aconstructive}.}. For $\ell$ layers, this yields  
\begin{equation}
N_{s}^{FALQON} \leq 2m\ell(d+1) = \mathcal{O}(md\ell).
\label{falqonsamp}
\end{equation}
This comparison suggests that FALQON has a more favorable sampling complexity than QAOA for cases where the number of QAOA optimization iterations $q(\ell)$ exceeds $d\ell$ in general, or $ d$ when a gradient algorithm is utilized.

\section{Combining FALQON and QAOA}
\label{FALQON+QAOA}

As we described previously, FALQON has flexibility in the choice of a control law, e.g., Eq.~\eqref{Eq:Discretebeta} and the value of $\beta_{1}$, the introduction of a reference perturbation (Sec.~\ref{Sec:RefPert}), and the choice of driver Hamiltonian. Once these features are selected, FALQON is a deterministic, constructive procedure, i.e., in the limit of perfect measurements, the resulting set of parameters $\{ \beta_{k} \}$ is uniquely specified for a problem Hamiltonian $\Hp$. On the other hand, QAOA has flexibility in the choice of the driver Hamiltonian(s), as well as the classical optimization method and \emph{all} initial values of the parameter set elements $\beta_{k}$ and $\gamma_{k}$.

The numerical results presented in Sec.~\ref{Sec:AppstoMaxCut} suggest that solving MaxCut using FALQON alone can require many layers and therefore may not be suitable for NISQ devices with limited circuit depths. In this section, we explore how FALQON results from a smaller number of layers can be used as a seed to initialize a QAOA circuit, thereby aiding in the subsequent search for optimal circuit parameters. Related work from Egger et al.~proposes a somewhat similar idea for ``warm-starting'' low-depth QAOA using the solution from a relaxation of the original combinatorial optimization problem \cite{Egger2021_Warm-starting}. While Sack and Serbyn introduce a ``Trotterized quantum annealing protocol'' to initialize QAOA \cite{Sack2021_Quantum}, parametrized by the time step $\Delta t$. In our work, we consider MaxCut on ensembles of unweighted 3-regular graphs with $n \in \{8, 10, 12, 14 \}$ vertices and 10-layer circuits implementing FALQON and QAOA. All of the simulations in this section are performed using \texttt{pyQAOA}, a Python-based simulator of QAOA circuits \cite{GvWinckel2021_pyQAOA}. As before, for graphs with $n = 8, 10$ vertices, we consider all nonisomorphic, connected 3-regular graphs. For graphs with $n = 12$ or $n = 14$ vertices, we consider a set of 50 randomly-generated nonisomorphic graphs for each value of $n$.

For each graph, the set of parameters $\{ \beta_{k} \}$ is generated for a 10-layer circuit using FALQON as described in Algorithm \ref{algo1} with $\Hd$ as specified in Eq.~\eqref{eq:sigmaxdriver}. To use the results of FALQON to initialize QAOA, the product $\beta_{k} \Delta t$ obtained from FALQON becomes the initial value for $\beta_{k}$ in QAOA; analogously, $\Delta t$ from FALQON becomes the initial value of \emph{all} $\gamma_{k}$ in QAOA, i.e.,
\begin{equation}
\beta_{k} \Delta t \rightarrow \beta_{k} \ \ \textrm{and} \ \ \Delta t \rightarrow \gamma_{k} \ \ \textrm{for all} \ k.
\label{eq:FALQON-to-QAOA}
\end{equation}
This follows from the relationship between FALQON unitary operations and QAOA unitary operations, i.e., compare $\Up$ and $\Ud(\beta_{k})$ in Eqs.~\eqref{eq:driverunitary}--\eqref{eq:problemunitary} to the corresponding QAOA operations.

Within the context of a 10-layer QAOA circuit, the two sets of parameters $\{ \beta_{k} \}$ and $\{ \gamma_{k} \}$ are then optimized using a quasi-Newton optimization method (BFGS). For simplicity, we refer to this sequential FALQON + QAOA procedure as ``FALQON+''. In addition, we compare the performance of FALQON+ to QAOA with multiple random initializations, i.e., we perform QAOA using multistart BFGS with 20 randomly-selected initial values for $\{ \beta_{k} \}$ and $\{ \gamma_{k} \}$.

\begin{figure*}[t]
\includegraphics[width=2.0\columnwidth]{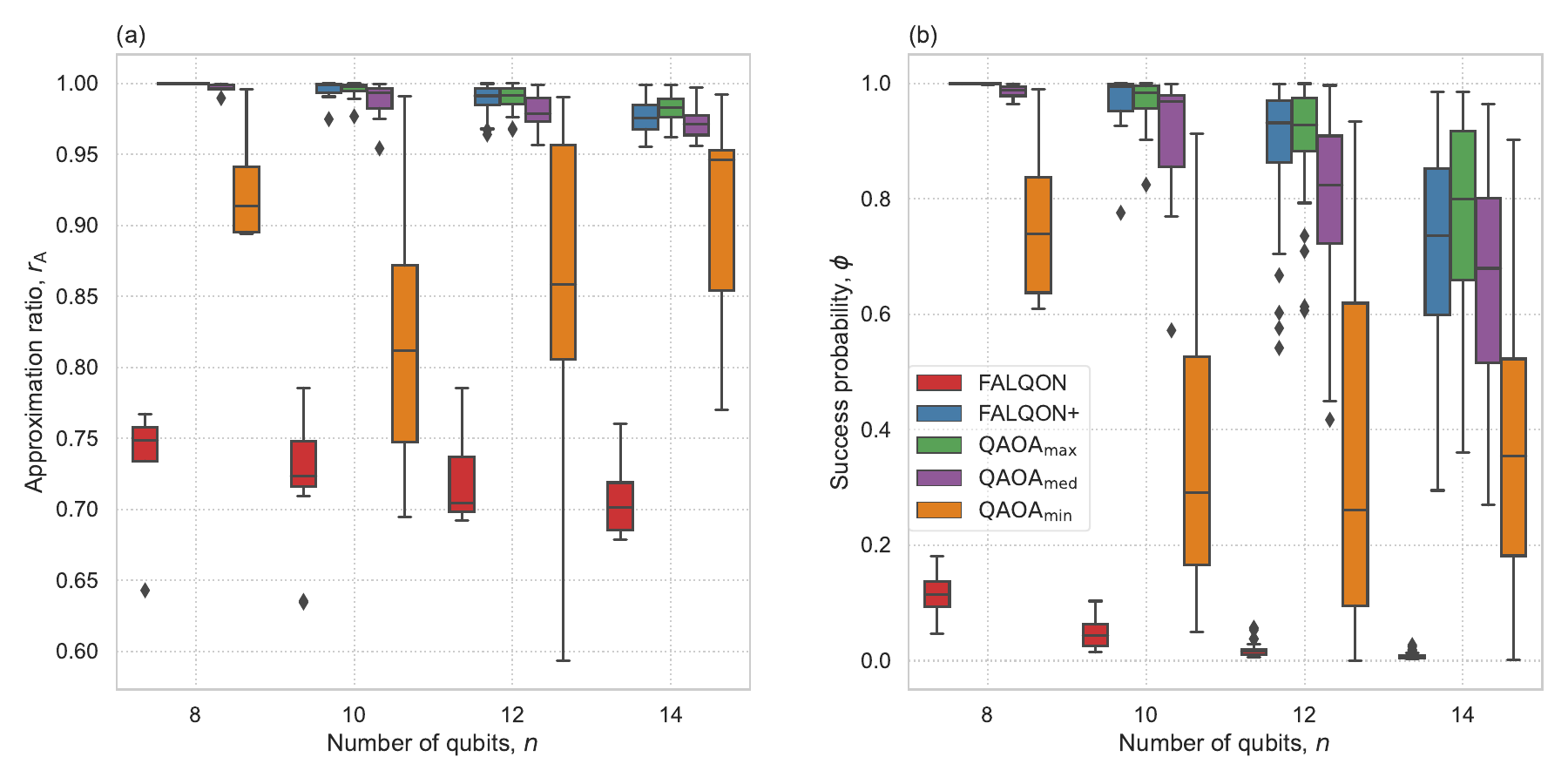}
\caption{Performance of MaxCut on ensembles of 3-regular graphs with $n$ vertices for $n \in \{8, 10, 12, 14 \}$ for FALQON, FALQON+, and multistart QAOA for 10-layer circuits, quantified by (a) the approximation ratio $r_{\textrm{A}}$ and (b) the success probability $\phi$, where $\mathrm{QAOA}_{\max}$, $\mathrm{QAOA}_{\mathrm{med}}$, and $\mathrm{QAOA}_{\min}$ denote distributions of maximum, median, and minimum QAOA results, respectively. In both figures, the color-shaded boxes and encompassed horizontal line represent the interquartile range and the median value of the data, respectively, while the whiskers extend out to 1.5 of the interquartile range; points beyond this range are identified as ``outliers'' (represented as grey diamond symbols).}
\label{fig:FALQON+ratios+state}
\end{figure*}

Approximation ratios $r_{\mathrm{A}}$ and success probabilities $\phi$ for FALQON, corresponding FALQON+, and multistart QAOA are reported in Fig.~\ref{fig:FALQON+ratios+state}. In both figures, the color-shaded boxes and encompassed horizontal line represent the interquartile range and the median value of the data, respectively, while the whiskers extend out to 1.5 of the interquartile range; points beyond this range are identified as ``outliers'' (represented as grey diamond symbols). Comparing the results of FALQON and FALQON+, note that in addition to improved approximation ratios, FALQON+ also substantially improves the success probabilities, at the cost of only one application of QAOA with BFGS. Since the final state measured at the end of the circuit corresponds to an actual (approximate) solution of the MaxCut problem, increasing success probabilities is more important than increasing approximation ratios. For our multistart QAOA simulations, we present distributions of maximum, median, and minimum (with respect to the randomly-selected then optimized sets of initial parameters $\{ \beta_{k} \}$ and $\{ \gamma_{k} \}$) approximation ratios and corresponding success probabilities for each ensemble of graphs. In the legend of Fig.~\ref{fig:FALQON+ratios+state}, these distributions are denoted as $\mathrm{QAOA}_{\max}$, $\mathrm{QAOA}_{\mathrm{med}}$, and $\mathrm{QAOA}_{\min}$, respectively. In principle, QAOA can perform better than FALQON overall since QAOA parameters can be optimized globally \emph{and} these parameters include the additional set $\{ \gamma_{k} \}$. However, in practice, achieving this improved performance may require multiple optimizations. For our simulations, FALQON+ performs comparably to the maximum and median cases of multistart QAOA.

\begin{figure}[t]
\includegraphics[width=1.0\columnwidth]{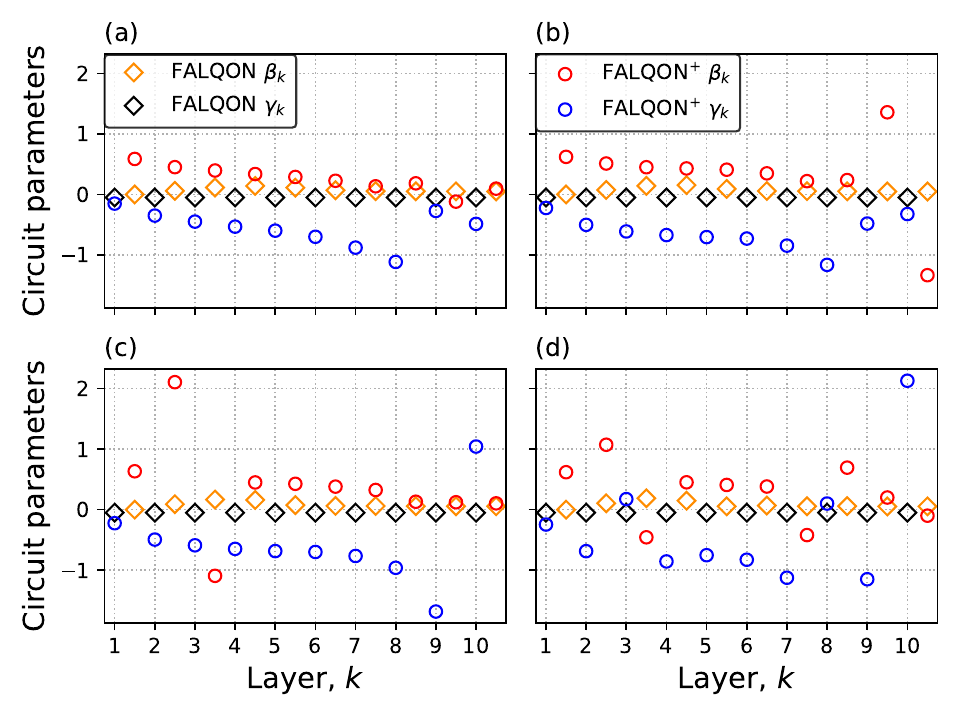}
\caption{Comparison of FALQON and corresponding FALQON+ parameters for example instances of 3-regular graphs with $n$ vertices for $n = 8$ (a), $n = 10$ (b), $n = 12$ (c), and $n = 14$ (d). Orange and black diamonds denote FALQON $\beta_{k} \Delta t$ and $\gamma_{k} = \Delta t$ parameters; red and blue circles denote FALQON+ $\beta_{k}$ and $\gamma_{k}$ parameters, as described in Eq.~\eqref{eq:FALQON-to-QAOA}. Because each FALQON and QAOA circuit layer contains a sequence of unitary operations, i.e., $\Ud(\beta_{k}) \Up(\gamma_{k})$ in Eqs.~\eqref{eq:QAOA-sequence} and \eqref{eq:FALQON-sequence}, we plot $\gamma_{k}$ at $k$ and $\beta_{k}$ at $k + 1/2$ for clarity.}
\label{fig:FALQON+theta}
\end{figure}

In Fig.~\ref{fig:FALQON+theta}, we present example instances of FALQON ($\beta_{k} \Delta t$ and $\Delta t$) and corresponding FALQON+ ($\beta_{k}$ and $\gamma_{k}$) parameters for $n \in \{8, 10, 12, 14 \}$ vertices \footnote{\label{fn:pyQAOA} In our version of \texttt{pyQAOA}, unitary operations are implemented as the conjugate transpose of Eqs.~\eqref{eq:driverunitary}--\eqref{eq:problemunitary}, i.e., $-i \rightarrow +i$ in the exponent. To account for this difference, we plot $-\beta_{k}$ and $-\gamma_{k}$ rather than $\beta_{k}$ and $\gamma_{k}$ in Fig.~\ref{fig:FALQON+theta}, hence the negative values for the FALQON circuit parameters $\gamma_{k}$, corresponding to $-\Delta t = -0.025$.}. Combined with results presented in Fig.~\ref{fig:FALQON+ratios+state}, these examples illustrate that substantial changes and improvements can occur between FALQON initialization and subsequent FALQON+ convergence, indicating that the parameters generated by FALQON, i.e., $\beta_{k} \Delta t$ and $\Delta t$, do not correspond to local optima for the QAOA landscape. In this sense, FALQON can prepare parameters for a successful application of QAOA, thereby reducing the expense of the optimization effort for QAOA. Although not presented here, the FALQON-QAOA parameter differences presented in Fig.~\ref{fig:FALQON+theta} are typical of all of our simulation results.

Based on results and analysis presented here, FALQON+ may provide a tractable solution to the challenge of identifying optimal parameters for QAOA. Overall, our results demonstrate that FALQON can be used to enhance the performance of depth-limited QAOA, with minimal additional cost. See ref.~\cite{magann2021feedbackbased} for estimates of FALQON and QAOA sampling complexity.

\section{Quantum annealing applications} \label{sec:annealing}

Quantum annealing \cite{das2008colloquium} is an approach for preparing the ground state of a problem Hamiltonian $\Hp$ that proceeds by initializing a quantum system in the ground state of another Hamiltonian $\Hd$, and then evolving the system via the time-dependent Hamiltonian 
\begin{equation}
H(t) = u(t)\Hd+(1-u(t)) \Hp
\end{equation}
for $t \in [0, T]$, where $u(t)$ is the quantum annealing schedule, with $u(0) = 1$ and $u(T) = 0$. Without known structure in $\Hp$ to exploit, often the simplest annealing schedule is linear, where $u(t) = 1 - t/T$, and we consider this in the following. Then, the aim is to choose $T$ to be large enough such that the system remains in the instantaneous ground state of $H(t)$ at all times, so that as $\Hp$ is slowly turned on, this will evolve the system into the ground state of $\Hp$ at time $T$. 

\begin{figure}[t]
\includegraphics[width=1.0\columnwidth]{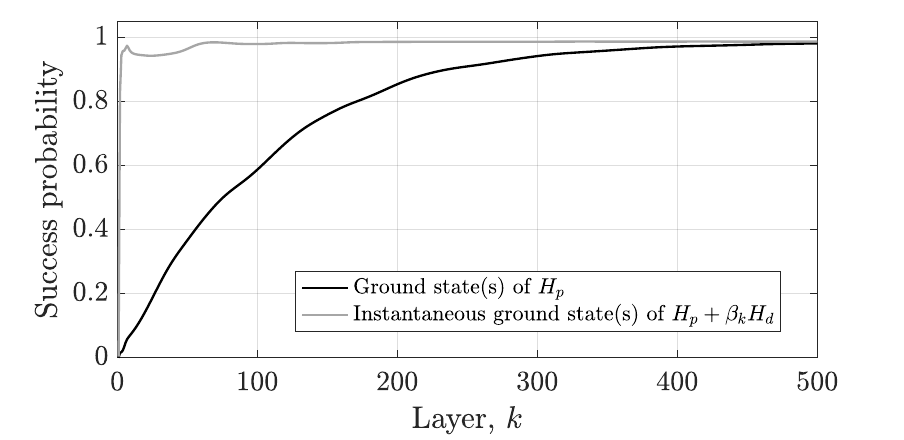}
\caption{The population in the ground state, $\phi$, and the population in the instantaneous ground state, $\phi_{\mathrm{inst}}$, are plotted as a function of layer, for an application of FALQON to MaxCut on an unweighted 3-regular graph with $n=8$ vertices. }
\label{Fig:InstGSOv}
\end{figure}

\begin{figure}
\includegraphics[width=1.0\columnwidth]{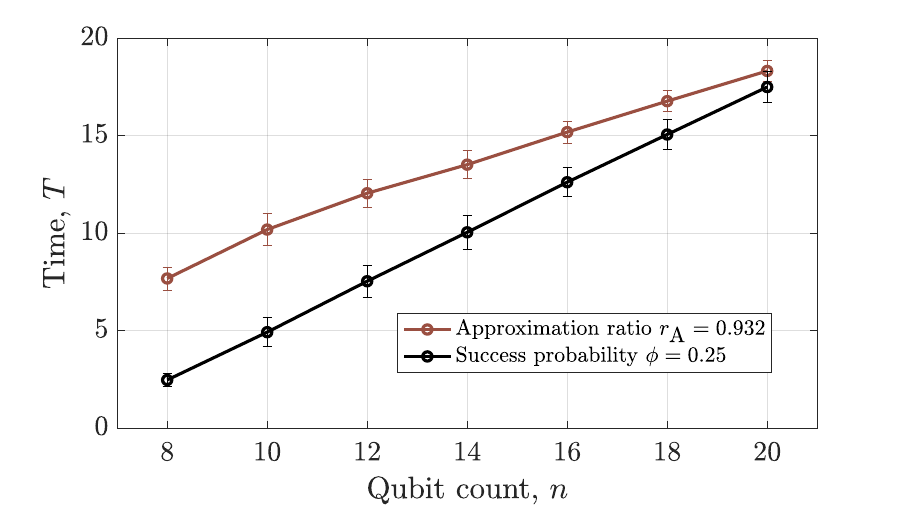}
\caption{The mean digitized time $T = 2k\Delta t$ needed to achieve the reference values of $r_{\textrm{A}} = 0.932$ (brown) and $\phi = 0.25$ (black) using a FALQON-inspired annealing schedule is shown for the unweighted MaxCut problems considered in Fig.~\ref{PRAMainResults}. The mean is taken over the set of unweighted, 3-regular graphs that are considered at each problem size $n$, and the error bars show the associated standard deviations.}
\label{Tresults}
\end{figure}

In this section, we compare FALQON to linear quantum annealing because of numerical evidence suggesting that FALQON may proceed via a similar adiabatic mechanism, i.e., by slowly switching on the problem Hamiltonian $\Hp$, such that the system remains in the instantaneous ground state. Evidence of this potential adiabatic behavior is shown in Fig.~\ref{Fig:InstGSOv}, for a representative instance of unweighted 3-regular MaxCut on $n=8$ vertices with $\Delta t = \Delta t_c$. The population in the instantaneous ground state, $\phi_{\mathrm{inst}}$, is computed as $\phi_{\mathrm{inst}} = \sum_{j} |\langle \psi| \tilde{q}_{0,j} \rangle|^2 $, where the sum is taken over $j$ degenerate instantaneous ground states (found numerically, as the eigenstates whose associated eigenvalues are within 0.01 of the lowest eigenvalue); the set of instantaneous eigenstates $\{\tilde{q}\}$ is computed by numerically diagonalizing $\Hp + \beta \Hd$ at each layer. The consistently high values of $\phi_{\mathrm{inst}}$ in Fig.~\ref{Fig:InstGSOv}, which is representative of the behavior seen across other MaxCut instances, suggest that FALQON may give rise to a Trotterized version of an adiabatic process, i.e., in which strong rotations are applied initially to transfer $|\psi\rangle$ into the instantaneous ground state, and then, the system remains primarily in the instantaneous ground state for the remaining evolution. In order to achieve this behavior, we see in Fig.~\ref{PRAMainResults}(c) that $\beta$ initially has large values, then decreases monotonically as a function of layer, similar to the behavior of an annealing schedule $u(t)$. Particularly notable from Fig.~\ref{PRAMainResults}(c) is that the $\beta$ curves appear to concentrate around a single average curve for each value of $n$, indicating that there may be a universal FALQON solution for this class of problems. As such, we relate these curves to digitized quantum annealing schedules, and consider the digitized time $T = 2k\Delta t$ needed to achieve $r_{\textrm{A}} = 0.932$ or $\phi = 0.25$ using FALQON. The results are shown in Fig.~\ref{Tresults}, which shows that $T$ scales favorably with respect to $n$, with an a linear scaling at the problem sizes evaluated.

\begin{figure}
\includegraphics[width=1\columnwidth]{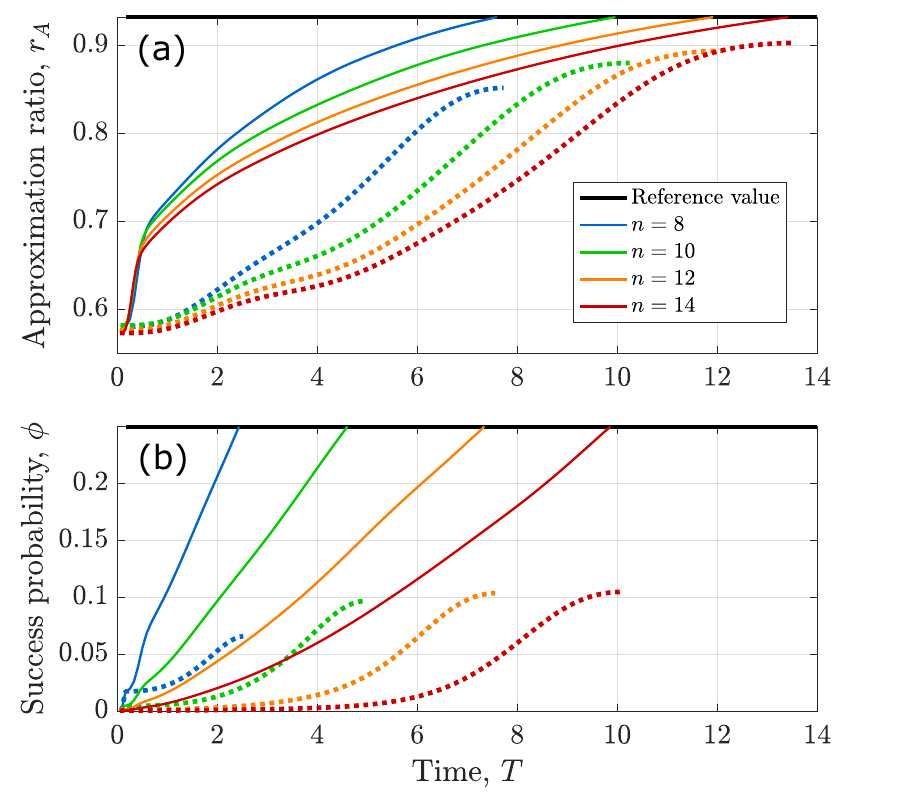}
\caption{(a) The mean approximation ratios, $r_A$, obtained by FALQON (solid curves) and a linear quantum annealing schedule (dotted curves) for MaxCut on unweighted, 3-regular graphs with $n=8,10,12,14$ vertices. The mean is taken over the set of different graphs that are considered at each problem size $n$. For the linear quantum annealing schedule, $T$ is chosen to be the time when FALQON reaches the reference value $r_{\textrm{A}}= 0.932$ for each value of $n$. This reference value is plotted as a horizontal black line. Panel (b) shows the corresponding results for the success probabilities, $\phi$. }
\label{Annealing}
\end{figure}

We then compare the performance of FALQON against that of a digitized linear quantum annealing schedule. We present the results of our numerical comparison in Fig.~\ref{Annealing} for the same MaxCut problem instances considered in Fig.~\ref{PRAMainResults}. Our findings indicate that for the same value of $T$, FALQON consistently shows stronger performance, as quantified by both $r_{\textrm{A}}$ and $\phi$.

It is of course important to note that this comparison is limited insofar as we compare only to a linear annealing schedule. This restriction was chosen for simplicity; it remains to be seen how FALQON compares relative to quantum annealing with various optimized schedules \cite{Roland_Cerf_2002, Brif_2014, Zeng_Zhang_Sarovar_2016}. Nevertheless, these results suggest that feedback-based protocols could be useful for improving performance of analog annealing devices as well. For example, the control schedule determined by FALQON, $\beta(t)$, could be used as the basis for an adiabatic annealing schedule. Alternatively, an annealing schedule could be derived through execution of an analogous Lyapunov-control inspired feedback strategy on an analog annealer, assuming the required measurements for determining $A(t)$ could be performed.

\section{Outlook}  

We have introduced FALQON as a constructive, feedback-based algorithm for solving combinatorial optimization problems using quantum computers, and explored its utility towards the MaxCut problem via a series of numerical experiments. Crucially, FALQON does not require classical optimization, unlike other quantum optimization frameworks such as QAOA. However, this advantage comes at a cost. As we found in our numerical illustrations, the quantum circuits needed tend to be much deeper than those conventionally considered in QAOA, suggesting that there is a tradeoff between the classical and quantum costs. 

Our numerical demonstrations utilized the feedback law given in Eq.~(\ref{Eq:Discretebeta}), although a much broader class of functions could be considered, as per Eq.~(\ref{Eq:betaQLC}), and the performance for different choices of $w$ and $f$ should be explored. Furthermore, the use of bang-bang control laws, e.g., where $\beta \in \{\pm \beta_{\max}\}$ switches between $\pm \beta_{\max}$, for a value of $\beta_{\max}$ chosen according to the sign of $A$, could also be considered in the future. Furthermore, we remark that the performance of FALQON depends on the choice of $\Delta t$, suggesting that it may be possible to design methods to optimally or adaptively choose $\Delta t$, e.g., for a given problem, or in a  layer-by-layer manner based on measurement data, perhaps informed by Eq. (\ref{eq:deltat}), in order to enhance the algorithm performance. 

In future implementations, FALQON could be used alone as a substitute for conventional QAOA, or it could be used in combination with QAOA (e.g., by taking $|\psi_0\rangle$ to be the terminal state from an already-optimized QAOA circuit). Similarly, in this work we have explored how FALQON could be used to seed QAOA, by identifying a set of initial QAOA parameters that can serve as the starting point for subsequent classical optimization. We expect that this seeding procedure may have particular benefit in settings with limited circuit depth, in cases where FALQON fails to converge on its own, and in cases where QAOA fails to converge on its own due to difficulty with effective initialization of the classical optimization procedure. 

We further remark that in situations where circuit depth is limited, FALQON can be extended to incorporate additional driver Hamiltonians, drawing on the framework outlined in Sec.~\ref{Sec:MultiHd}, and it could also be modified to use a hardware-inspired ansatz, where the circuit is formed by alternating rounds of an Ising Hamiltonian defined by the hardware connectivity, denoted as $H_{\textrm{h}}$, and a driver Hamiltonian $\Hd$, while the objective remains determined by the Ising problem Hamiltonian, denoted as $\Hp$. In this scenario, despite changes in the structure of the quantum circuits, the measurements of $ i[\Hd, \Hp]$ needed to assign values to the $\beta$ parameters would remain unchanged.

\acknowledgments

We acknowledge discussions with C. Arenz, L. Brady, L. Cincio, T.S. Ho, L. Kocia, O. Parekh, H. Rabitz, and K. Young. MDG gratefully acknowledges helpful discussions about \texttt{pyQAOA} usage with G. von Winckel \cite{GvWinckel2021_pyQAOA}. This work was supported by the U.S. Department of Energy, Office of Science, Office of Advanced Scientific Computing Research, under the Quantum Computing Application Teams program. A.B.M. also acknowledges support from the U.S. Department of Energy, Office of Science, Office of Advanced Scientific Computing Research, Department of Energy Computational Science Graduate Fellowship under Award Number DE-FG02-97ER25308, as well as support from Sandia National Laboratories’ Laboratory Directed Research and Development Program. M.D.G. also acknowledges support from the U.S. Department of Energy, Office of Science, Advanced Scientific Computing Research, under the Accelerated Research in Quantum Computing (ARQC) program. SAND2022-14512 J.

This article has been authored by an employee of National Technology \& Engineering Solutions of Sandia, LLC under Contract No. DE-NA0003525 with the U.S. Department of Energy (DOE). The employee owns all right, title and interest in and to the article and is solely responsible for its contents. The United States Government retains and the publisher, by accepting the article for publication, acknowledges that the United States Government retains a non-exclusive, paid-up, irrevocable, world-wide license to publish or reproduce the published form of this article or allow others to do so, for United States Government purposes. The DOE will provide public access to these results of federally sponsored research in accordance with the DOE Public Access Plan {https://www.energy.gov/downloads/doe-public-access-plan}. This paper describes objective technical results and analysis. Any subjective views or opinions that might be expressed in the paper do not necessarily represent the views of the U.S. Department of Energy or the United States Government.

This report was prepared as an account of work sponsored by an agency of the United States Government. Neither the United States Government nor any agency thereof, nor any of their employees, makes any warranty, express or implied, or assumes any legal liability or responsibility for the accuracy, completeness, or usefulness of any information, apparatus, product, or process disclosed, or represents that its use would not infringe privately owned rights. Reference herein to any specific commercial product, process, or service by trade name, trademark, manufacturer, or otherwise does not necessarily constitute or imply its endorsement, recommendation, or favoring by the United States Government or any agency thereof. The views and opinions of authors expressed herein do not necessarily state or reflect those of the United States Government or any agency thereof. 

\bibliography{bib}

\newpage
\appendix

\section{Convergence of QLC}
\label{AppA:Convergence}

Within the QLC framework outlined in Sec.~\ref{Sec:QLC}, it has been shown that asymptotic convergence to the ground state of $\Hp$ can be guaranteed when the following sufficient criteria are met \cite{1272601, lyapunovsurvey, BEAUCHARD2007388, doi:10.1002/rnc.1748}:
\begin{enumerate}
    \item $\Hp$ has no degenerate eigenvalues, i.e., $q_i\neq q_j$ for $i\neq j$ where $q_i$ and $q_j$ are the $i$-th and $j$-th eigenvalues of $\Hp$
    \item $\Hp$ has no degenerate eigenvalue gaps, i.e., $\omega_{ji}\neq \omega_{lk}$ for $(i,j)\neq(k,l)$, where $\omega_{ji}=q_j-q_i$ is the gap between the $i$-th and $j$-th eigenvalues of $\Hp$
    \item  $\langle q_j|\Hd|q_i\rangle\neq 0$ for all $i\neq j$ 
    \item  $\Ep(|q_0\rangle)<\Ep(|\psi(t=0)\rangle)<\Ep(|q_1\rangle)$ 
\end{enumerate}
In particular, if criteria 1-3 are met, then the LaSalle invariance principle \cite{la1976stability} can be used to show that any initial state $|\psi(t=0)\rangle$ will converge asymptotically to the largest invariant set, i.e., the largest set of states where $\frac{d}{dt}\Ep = 0$. When $\Ep$ is chosen per Eq.~(\ref{Eq:DefineLFunc}), it can be shown that the largest invariant set is the set of eigenstates of $\Hp$. Within this set, the eigenstate $|q_0\rangle$ with the smallest eigenvalue is the minimum, the eigenstate with the largest eigenvalue is the maximum, and all other eigenstates with intermediate eigenvalues are saddle points. In order to ensure convergence to the desired critical point $|q_0\rangle$, criterion (4) stipulates that the value of $\Ep(|\psi(0)\rangle)$ at time $t=0$ must be strictly lower than $\Ep(|q_1\rangle)=q_1$, such that the only critical point inside $\Ep(|\psi(t)\rangle) \leq \Ep(|\psi(0)\rangle)$ is the desired target, $|q_0\rangle$. Thus, the satisfaction of criteria (1)-(4) is sufficient to ensure that the system state will converge asymptotically to the desired target $|q_0\rangle$ \cite{1272601, lyapunovsurvey, BEAUCHARD2007388, doi:10.1002/rnc.1748}.

\section{Selecting $\Delta t>\Delta t_c$}
\label{AppA}

\begin{figure}[h!]
\includegraphics[width=1.0\columnwidth]{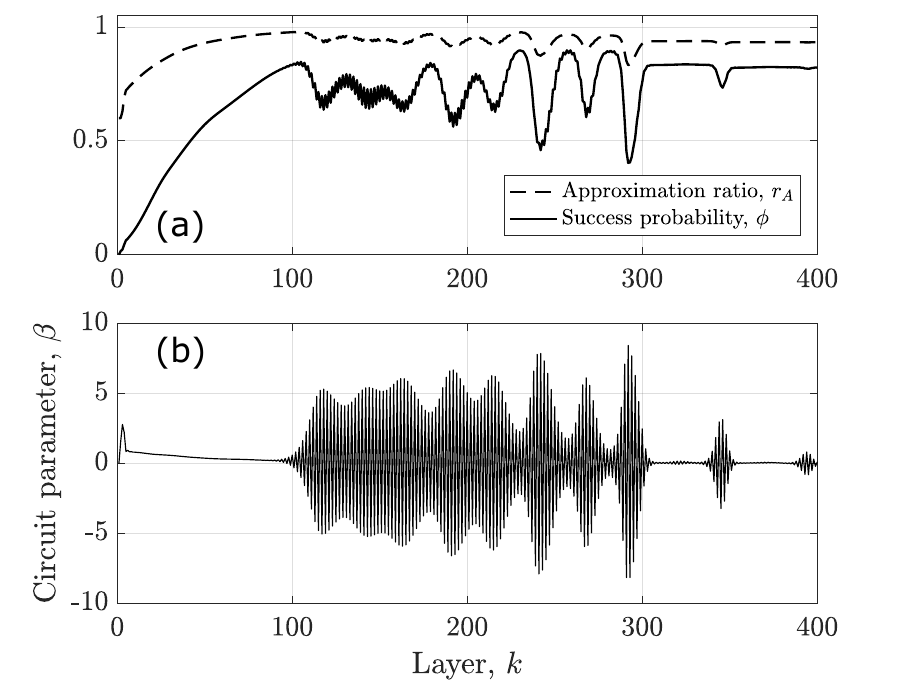}
\caption{Typical behavior when the time step $\Delta t$ is chosen to be too large, leading to a violation of the QLC criterion that $\langle \Hp\rangle$ decreases monotonically with respect to layer, $k$. In (a), the behavior of the approximation ratio, $r_{\textrm{A}}$, and the success probability, $\phi$, are shown, with the violation in monotonicity occurring around 100 layers. In (b), the associated behavior of the circuit parameter $\beta$ is plotted, indicating that the violation of the QLC criterion corresponds to the creation of rapid oscillations in $\beta$.  }
\label{Fig:Bigdt}
\end{figure}

Fig.~\ref{Fig:Bigdt} illustrates the effects of selecting a time step that is too large, i.e., $\Delta t > \Delta t_c$, for an instance of unweighted, 3-regular MaxCut on $n=8$ vertices, with $\Delta t = 0.065$. The behavior in Fig.~\ref{Fig:Bigdt} is representative of the behavior seen across other instances when $\Delta t > \Delta t_c$. In general, there is a balance between selecting a large $\Delta t$ for improving convergence \emph{and} satisfying $\Delta t \leq \Delta t_c$ for ensuring monotonic improvement in $\Ep$.

\begin{figure}[h!]
\includegraphics[width=1.0\columnwidth]{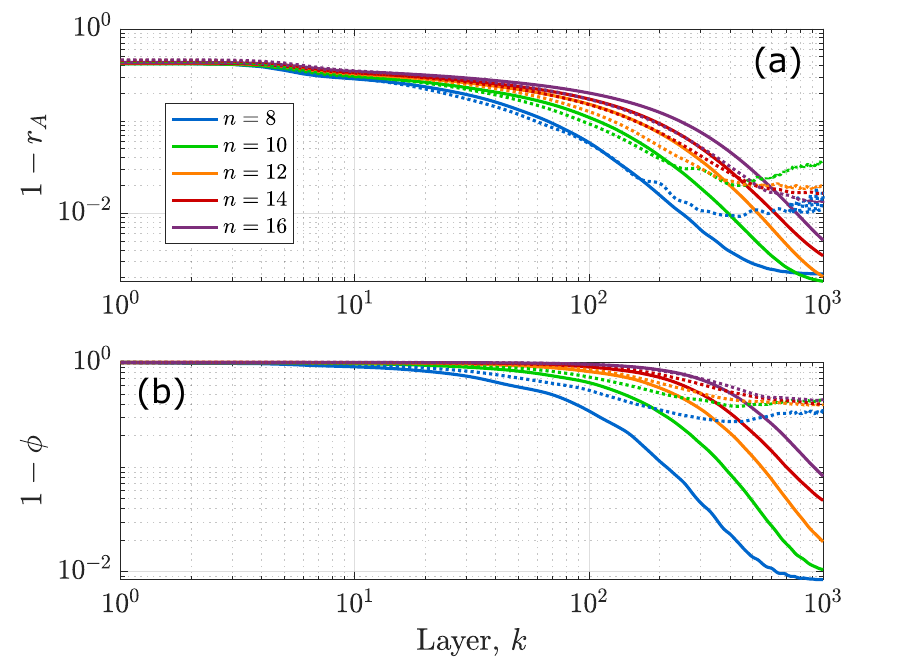}
\caption{The performance of FALQON, as quantified by the approximation ratio $r_{\textrm{A}}$ in (a) and the success probability $\phi$ in (b), is shown as a function of the layer $k$ for different qubit counts $n$ on a log-log scale. The same results are plotted on a linear scale in Fig.~\ref{PRAMainResults}. The solid curves show the average results for unweighted 3-regular graphs, and the dotted curves show average results for weighted 3-regular graphs. }
\label{Fig:Loglog}
\end{figure}

\section{Log-log plot of MaxCut results }
In Fig.~\ref{Fig:Loglog}, we plot the results presented in the main text in Fig.~\ref{PRAMainResults} using a log-log scale. 

\end{document}